\documentclass[10pt]{amsart}  
\usepackage{amssymb,amsmath,amscd,verbatim, mathrsfs}
\usepackage[all]{xy} 
\usepackage{color}

\font\tenrsf=rsfs10 at 11pt 
\font\sevenrsf=rsfs7 at 8pt 
\font\fiversf=rsfs5 at 6pt 
\newfam\rsffam 
\textfont\rsffam=\tenrsf 
\scriptfont\rsffam=\sevenrsf 
\scriptscriptfont\rsffam=\fiversf 
\def\rond#1{{\tenrsf\fam\rsffam#1}} 

\renewcommand{\theequation}{\arabic{section}.\arabic{equation}}
\newtheorem{theorem}{Theorem}[section]
\newtheorem{lemma}[theorem]{Lemma}
\newtheorem{proposition}[theorem]{Proposition}

\theoremstyle{definition} 
\newtheorem{definition}[theorem]{Definition}

\theoremstyle{remark} 
\newtheorem{rem}[theorem]{Remark} 

\newcommand{\Br}{\rond{B}}

\newcommand{\Dr}{\rond{D}} 
\newcommand{\Gr}{\rond{G}} 
\newcommand{\Hr}{\rond{H}}

\newcommand{\Kr}{\rond{K}} 
 
\newcommand{\rP}{\rond{P}}

\newcommand{\Bc}{\mathcal{B}} 
\newcommand{\Cc}{\mathcal{C}} 
\newcommand{\Dc}{\mathcal{D}}

\newcommand{\Ic}{\mathcal{I}} 
\newcommand{\Jc}{\mathcal{J}}

\newcommand{\im}{\mathrm{Im}}  
\newcommand{\id}{\mathrm{Id}}
\newcommand{\un}{\mathrm{1}}

\newcommand{\C}{\mathbb{C}} 
\newcommand{\R}{\mathbb{R}} 
\newcommand{\N}{\mathbb{N}}

\newcommand{\dG}{{\rm d}\Gamma}
\newcommand{\rv}{{\rm v}}
\newcommand{\slim}{\mathrm{s-}\!\lim} 
\newcommand{\wlim}{\mathrm{w-}\!\lim} 
\font\teneuf=eufm10 at 12pt
\font\seveneuf=eufm7 at 8pt
\font\fiveeuf=eufm5 at 6pt
\newfam\euffam
\textfont\euffam=\teneuf
\scriptfont\euffam=\seveneuf
\scriptscriptfont\euffam=\fiveeuf
\def\goth#1{{\teneuf\fam\euffam#1}}
\newcommand{\hg}{\goth{h}}
\def\build#1_#2^#3{\mathrel{\mathop{\kern 0pt#1}\limits_{#2}^{#3}}}
\def\cchi{\raisebox{.45 ex}{$\chi$}}
\def\pp{<}  
\def\pg{>} 

\begin{document} 
\title[On the Mourre estimate]
{Positive commutators, Fermi golden rule and the spectrum of zero
temperature Pauli-Fierz Hamiltonians. }
\author{Sylvain Gol\'enia}  
\address{Mathematisches Institut der Universit\"at Erlangen-N\"urnberg
Bismarckstr.\ 1 1/2 \\
91054 Erlangen, Germany}
\email{golenia@mi.uni-erlangen.de} 
\begin{abstract}  
We perform the spectral analysis of a zero temperature Pauli-Fierz
system for small coupling constants. Under the hypothesis of Fermi golden
rule, we show that the embedded eigenvalues of the uncoupled system
disappear and establish a limiting absorption principle above this level of
energy. We rely on a positive commutator approach introduced by
Skibsted and pursued by Georgescu-G\'erard-M\o ller. We complete some
results obtained so far by Derezi\'nski-Jak\u si\' c on one side and by
Bach-Fr\"ohlich-Segal-Soffer on the other side. 
\end{abstract}  
\maketitle  
\begin{center}
\emph{En hommage au 60\`eme anniversaire de Vladimir Georgescu.}
\end{center}
\section{Introduction}
\setcounter{equation}{0}
Pauli-Fierz operators are often used in quantum physics as generator
of approximate dynamics of a (small) quantum system interacting with a free
Bose gas. They describe typically a non-relativistic atom interacting
with a field of massless scalar bosons. 
Pauli-Fierz operators appear also in solid state physics. They are
used to describe the  interaction of phonons with a quantum system
with finitely many degrees of freedom. This paper is devoted to the
justification of the second-order perturbation theory  for a large
class of perturbation. For positive temperature system, this property
is related to the return to equilibrium, see for instance
\cite{DJ2} and reference therein. 

This question has been studied in many places,
see for instance  \cite{BFS, BFS2, BFSS, DJ, FMS, FP, HHH} for zero temperature
systems and \cite{DJ, JP, Merkli} for positive temperature. We mention also 
\cite{FGS, GGM, hsp, Ski} who studied certain spectral properties 
using positive commutator techniques. Here, we focus on the
zero temperature setting.  In \cite{BFS}, one initiates the analysis
using analytic deformation techniques. In \cite{BFSS} and in
\cite{DJ}, one introduces some kind of Mourre estimate approach. In the
former, one enlarges the class of perturbation studied in \cite{BFS}
and in the latter, one introduces another class. These two classes do
not fully overlap. This   is due to the choice of the conjugate
operator. In this paper, we enlarge the class of perturbations used in
\cite{DJ} for the question of the Virial theorem (one-commutator
theory) and also for the limiting absorption principal (two-commutator
theory).

Now, we present the model. For the sake of simplicity
and as in \cite{DJ}, we start with a $n$-level atom. It is
described by a self-adjoint matrix $K$ acting on a finite dimensional
Hilbert space $\Kr$.  Let $(k_i)_{i=0,\ldots, n}$ be its eigenvalues,
with $k_i<k_{i+1}$. On the other hand,  we have the Bosonic field
$\Gamma_s(\hg)$ with the $1$-particle Hilbert space 
$\hg:=L^2(\R^d, dk)$. The Hamiltonian is given by the second
quantization $\dG(\omega)$ of $\omega$, where $\omega(k)=|k|$, see
Section \ref{s:bosonic}. This is a massless and zero temperature system.
The free operator is given by $H_0= K\otimes
\un_{\Gamma(\omega)} +\un_\Kr\! \otimes \dG(\omega)$ on  $\Kr\!\otimes
\Gamma(\hg)$. Its spectrum is $[k_0, \infty)$. It has no singularly
continuous spectrum. Its point spectrum is the same as $K$, with
the same multiplicity. Let $\alpha\in\Bc(\Kr, \Kr\!\otimes \hg)$ be a
form-factor and $\phi(\alpha)$ the field operator associated to it,
see Section \ref{s:inter}.  Under the condition 
\begin{enumerate}
\item[{\bf (I0)}] $(1\otimes \omega^{-1/2})\alpha\in \Bc(\Kr, \Kr\!\otimes\hg)$,
\end{enumerate}
we define the interacting Hamiltonian on $\Kr\!\otimes \Gamma(\hg)$ by  
\begin{eqnarray}\label{e:H}
H_\lambda := 	K\otimes \un_{\Gamma(\omega)} +\un_\Kr\! \otimes
\dG(\omega) + \lambda \phi(\alpha), \mbox{ where } \lambda\in \R. 
\end{eqnarray}
The operator is self-adjoint with domain $\Kr\!\otimes \Dc\big(
\dG(\omega)\big)$. 

We now focus on a selected eigenvalue $k_{i_0}$, with $i_0>0$. The aim
of this paper is to give hypotheses on the form factor  $\alpha$ to
ensure that $H_\lambda$ has no eigenvalue in a neighborhood of
$k_{i_0}$ for $\lambda$ small enough (and non-zero). First, we have to
ensure that the perturbation given by the field operator will really
couple the system at energy $k_{i_0}$; we have to avoid form factors
like $\alpha(x)= \un \otimes b$ for all $x\in \Kr$ and some 
$b\in \hg$, see Section \ref{s:thre}. Here
comes the second-order perturbation theory,   namely the hypothesis of 
\emph{Fermi golden rule} for the couple $(H_0, \alpha)$ at energy $k_{i_0}$: 
\begin{eqnarray}\label{e:FGR0'}
\wlim_{\varepsilon \rightarrow 0^+}
P\phi(\alpha)\overline{P}\,\im(H_0-k +i\varepsilon)^{-1}
\overline{P}\phi(\alpha) P >0, \mbox{ on } P\Hr, 
\end{eqnarray} 
where $P:=P_{k_{i_0}}\otimes P_\Omega$ and $\overline P:=1-P$. 
At first sight, this is pretty implicit. 
We make it explicit in Appendix \ref{s:level}. This condition involves
the form factor, the eigenvalues of $H_0$ lower than $k_{i_0}$ and its
eigenfunctions. Therein, we also explain why the ground state energy
is tacitly excluded. 

In this paper, we are establishing an extended Mourre estimate, in the
spirit  of \cite{GGM2, Ski}; this is an extended version of the
positive commutator  technique initiated by E.\ Mourre, see
\cite{ABG,M} and \cite{G2, GJe} for recent developments.  Due to the
method, we make further hypotheses on the form-factor.  To formulate
them, we shall take advantage of the \emph{polar   coordinates} and of
the  unitary map:   
\begin{eqnarray}\label{e:T}
\quad T:=\left\{\begin{array}{cll}
	L^2(\R^d, dk)&\longrightarrow& L^2(\R^+, dr)\otimes
	L^2(S^{d-1}, d\theta) :=\tilde \hg  
	\\
	u&\longmapsto& Tu:= (r, \theta)\mapsto r^{(d-1)/2}u(r\theta).
\end{array}\right.
\end{eqnarray}
We identify $\hg$ and $\tilde \hg$ through this transformation. We
write $\partial_r$ for $\partial_r\otimes \un$. We first give meaning
to the commutator via: 
\begin{enumerate}
\item[{\bf (I1a)}] $\alpha\in \Bc\big(\Kr, \Kr\!\otimes
  \dot\Hr^1(\R^+)\otimes L^2(S^{d-1})\big)$,  
$1\otimes \omega^{-1/2} \partial_r \alpha \in \Bc(\Kr,  \Kr\!\otimes\hg)$.
\end{enumerate}
Here, the dot means the completion of $\Cc^\infty_c(\R^+)$ under the
norm given by the space. We denote by $\|\cdot\|_2$ the $L^2$ norm. 
Recall the norm of $\Hr^1$ is given by  $\|\cdot \|_2 + \|\partial_r\cdot\|_2$.

We explain the method on a formal level. We start by choosing a
conjugate operator so as to obtain some positivity of the
commutator. We choose  $A:=1_\Kr\!\otimes \dG(i\partial_r)$. Note this
operator is not self-adjoint and only maximal symmetric. We set
$N:=1_\Kr\!\otimes \dG(\id)$, the \emph{number operator}. Thanks to
{\bf (I1a)}, one obtains 
\begin{align*}
[H_\lambda, iA] = \underbrace{N + 1_\Kr\!\otimes P_\Omega}_{\geq 1} +
\underbrace{\lambda\phi(\partial_r \alpha) - 
1_\Kr\!\otimes P_\Omega}_{H_\lambda \mbox{-bounded }} =: M+ S.
\end{align*}
 Consider a compact interval $\Jc$. Since $\dG(\omega)$ is
 non-negative, we have:    
\begin{eqnarray}\label{e:H0p}
E_\Jc(H_0)= \sum_{0\leq i \leq 
\sup(\Jc)} P_{k_i}\otimes E_{\Jc-k_i}\big(\dG(\omega)\big).
\end{eqnarray}
We infer $(1_\Kr\!\otimes P_\Omega) E_\Jc(H_0)=0$ if and only if $\Jc$
contains no eigenvalues of $K$.  We evaluate the commutator at an
energy $\Jc$ which contains $k_{i_0}$ and no other $k_i$. Thus, 
\begin{align}\label{e:Ipositiv}
M+ E_\Jc(H_0) SE_\Jc(H_0) \geq 1+ \big(-1+O(\lambda)\big)E_\Jc(H_0)
\geq O(\lambda)E_\Jc(H_0), 
\end{align}
since $\phi(i\partial_r \alpha)$ is $H_0$-bounded. We keep $M$ outside
the spectral measure as it is  not $H_\lambda$-bounded. Note we have
no control on the sign of $O(\lambda)$ so far. We have not yet used
the Fermi golden rule assumption. We follow an idea of \cite{BFSS} and
set  
\begin{align*}
B_\varepsilon := \im
\big(\big((H_0-k_{i_0})^2+\varepsilon^2\big)^{-1}\overline{P}
\phi(\alpha)P\big ) 
\end{align*}
Observe that  \eqref{e:FGR0'} implies there exists $c>0$ such that 
\begin{align*}
P[H_\lambda, i \lambda B_\varepsilon]P= \frac{\lambda^2}{\varepsilon}
P \phi (\alpha)\overline{P} \, \im (H_0-k_{i_0}+ i \varepsilon)^{-1}\, 
\overline{P} \phi (\alpha) P \geq \frac{c\lambda^2}{\varepsilon}P, 
\end{align*}
holds true for $\varepsilon$ small enough. Let $\hat A:= A + \lambda
B_\varepsilon$  and $\hat S := S + \lambda [H_\lambda,i
  B_\varepsilon]$. We have $[H_\lambda, i\hat A] = M + \hat S$. We go
back to   \eqref{e:Ipositiv} and infer:
\begin{align}\label{e:Ipositiv2}
M+ E_\Jc(H_0) \hat S E_\Jc(H_0) \geq \big( c\lambda^2/\varepsilon +O(\lambda)\big) E_\Jc(H_0) + \mbox{ error terms}.
\end{align}
By taking $\varepsilon:= \varepsilon(\lambda)$, one hopes to obtain
the positivity of the constant in front  of $E_\Jc(H_0)$, to control
the errors terms and to replace the spectral measure by the one of
$H_\lambda$. Using the Feshbach method and with a more involved choice
of conjugate operator, we show in Section \ref{s:thre} that there are
$\lambda_0, c', \eta>0$ so that  
\begin{align}\label{e:Ipositiv3}
M+ E_\Jc(H_\lambda) \hat S E_\Jc(H_\lambda) \geq c'|\lambda|^{1+\eta}
E_\Jc(H_\lambda), \mbox{ for all } |\lambda|\leq \lambda_0, 
\end{align}
on the sense of forms on $\Dc(N^{1/2})$.

One would like to deduce there is no eigenvalue in $\Jc$ from
\eqref{e:Ipositiv3}.  To apply a Virial theorem, one has at least to
check that the eigenvalues of $H_\lambda$ are in the domain of
$N^{1/2}$. One may proceed like in  \cite{M}. In this article, we
follow \cite{GGM, Ski} and construct a sequence of approximated
conjugate operators $\hat A_n$ such that $[H_\lambda, i\hat A_n]$ is
$H_\lambda$-bounded, converges to $[H_\lambda, i\hat A]$  and such
that one may apply the Virial theorem with $A_n$. To justify these
steps,  we make a new assumption: 
\begin{enumerate}
\item[{\bf (I1b)}] $\un_\Kr\! \otimes \omega^{-a}\alpha\in \Bc(\Kr,
  \Kr\!\otimes \hg)$, \mbox{ for some } $a>1$. 
\end{enumerate}
We now give our first result, based on the Virial theorem, see
Proposition \ref{p:6.5}. 

\begin{theorem}\label{t:intro0}
Let $\Ic$ be an open interval containing $k_{i_0}$ and no other $k_i$. Assume
the Fermi golden rule hypothesis \eqref{e:FGR0'} at energy $k_{i_0}$.
Suppose that {\bf(I0)}, {\bf(I1a)} and {\bf (I1b)}  are
satisfied. Then, there is $\lambda_0>0$ such that $H_\lambda$ has no
eigenvalue in $\Ic$, for all  $|\lambda|\in (0, \lambda_0)$.
\end{theorem}

We now give more information on the resolvent $R_\lambda(z):=(H_\lambda
-z)^{-1}$ as the imaginary part of $z$ tends to $0$. We show it
extends to an operator in some weighted spaces around the real axis. 
This is a standard result in the Mourre theory, when one supposes some 
$2$-commutators-like hypothesis, see \cite{ABG}. Here, as the
commutator is not $H_\lambda$-bounded, one relies on an adapted
theory. We use \cite{GGM} which is a refined version of \cite{Ski}.  
We check  the hypotheses {\bf (M1)}--{\bf (M5)} given in Appendix
\ref{s:ms} and deduce a limiting absorption principle, thanks to
Theorem \ref{t:LAP}.  Using again \eqref{e:T}, we state our
 class of form factors: 
\begin{enumerate}
\item[{\bf (I2)}] $\alpha\in \Bc\big(\Kr, \Kr\!\otimes
 \dot\Br^{1,1}_2(\R^+)\otimes L^2(S^{d-1})\big)$.
\end{enumerate}
Recall that the dot denotes the completion of $\Cc^\infty_c$. One
choice of norm for $\Br^{1,1}_2$ is:  
\begin{eqnarray*}
\|f\|_{\Br^{1,1}_2(\R^+)}= \|f\|_2+
\int_0^1\big\|f(2t+\cdot)-2f(t+\cdot)+f(\cdot) \big\|_2\, \frac{dt}{t^2}. 
\end{eqnarray*} 
We refer to \cite{ABG, Tri} for Besov spaces and real
interpolation. To express the weights, consider $\tilde b$ the square root
of the Dirichlet Laplacian on $L^2(\R^+, dr)$. Using
\eqref{e:T}, we define $b:=\un_\Kr\!\otimes T^{-1}\tilde bT$ in $\Hr$. 
Set $\rP_s :=\un_\Kr\!\otimes (\dG(b)+1)^{-s}(N+1)^{1/2}$.
\begin{theorem}\label{t:intro1}
Let $\Ic$ be an open interval containing $k_{i_0}$ and no other $k_i$. Assume
the Fermi golden rule hypothesis \eqref{e:FGR0'} at energy $k_{i_0}$.
Suppose that {\bf(I0)}, {\bf(I1a)} and {\bf (I2)}
(and not necessarily  {\bf (I1b)}), there is $\lambda_0>0$ such that
$H_\lambda$ has no eigenvalue in $\Ic$, for all  $|\lambda|\in (0, \lambda_0)$.
Moreover, $H_\lambda$ has no singularly continuous spectrum in
$\Ic$. For each compact interval $\Jc$ included in $\Ic$, and for all
$s\in(1/2, 1]$, the limits  
\begin{align*}
\rP_s^* R_\lambda(x \pm i0)\,
 \rP_s:=\lim_{y\rightarrow 0^+} \rP_s^*R_\lambda(x \pm iy) \rP_s
\end{align*} 
exist in norm uniformly in $x\in\Jc$. Moreover the maps:
\begin{align*}
\Jc\ni x\mapsto \rP_s^* R_\lambda(x \pm i0)  \rP_s
\end{align*}
are H\"older continuous of order $s-1/2$ for the norm topology of $\Bc(\Hr)$
\end{theorem}

To our knowledge, the condition {\bf (I2)} is new, even for the
question far from the thresholds. We believe it to be optimal in the Besov scale for
limiting absorption principle.

We now compare our result with the literature. In \cite{BFSS}, they
use a different conjugate operator, the second quantization
of the generator of dilatation. With this choice they have $[H_0, iA]=
\un_\Kr\!\otimes \dG(\omega)$. The commutator is
$H_\lambda$-bounded. They modify  the conjugate operator in the same way
but the choice of parameters is more involved. The class of
perturbations is thus different from ours. 

In \cite{DJ}[Theorem 6.3], one shows the absence of embedded eigenvalues
by proving a limiting  absorption principal with
the weights $\un_\Kr\!\otimes (\dG(b)+1)^{-s}$, for $s>1/2$, without
any contribution in $N$.  
They suppose essentially {\bf (I0)} and that $\alpha\in \Bc\big(\Kr,
\Kr\!\otimes \dot\Hr^s(\R^+)\otimes L^2(S^{d-1})\big)$, for $s>1$.
The class of perturbations is chosen in relation with the weights. 
Their strategy is to take advantage the Fermi golden rule at the level of the
limiting absorption principle, with the help of the Feshbach
method. The drawback is that they are  limited by the relation
weights/class of form-factors and they cannot give a  Virial-type
theorem. On the other hand, their method allows to cover some positive
temperature systems and we do not deal with this question. Their
method leads to fewer problems with domains questions. We mention that
they do not suppose the second condition of {\bf (I1a)}. 

Therefore, concerning merely the disappearance of the eigenvalues,
the conditions  {\bf (I1a)} and {\bf (I1b)} do not imply $\alpha$ 
to be better than $\dot\Hr^1(\R^+)$, in the Sobolev
scale. Hence, Theorem \ref{t:intro0} is a new result. We point out
that the condition {\bf (I2)} is weaker than the one used
in \cite{DJ}.  The weights obtained in the limiting absorption principle are also better
than the ones given in \cite{DJ}. We mention that one could improve
them by using some Besov spaces, see \cite{GGM}.  To
simplify the presentation, we do not present them here. We believe
they could hardly be reached by the method exposed in \cite{DJ} due to
the interplay between weights and form-factors. 

In \cite{GGM2} and in \cite{Ski}, one cares about showing that the
point spectrum is locally finite, i.e.\ without clusters and of
finite multiplicity. Here, they use a Virial theorem. Between the
eigenvalues, one shows a limiting absorption principle, and uses a
hypothesis on the second commutator, 
something stronger than {\bf (I2)}, see Section \ref{s:2comm}. In our
approach, we use the Virial theorem and the  limiting
absorption principle in an independent way. In particular, if  
one is interested only in the limiting absorption principle, one does
not need to suppose the more restrictive condition {\bf (I1b)} but only {\bf
  (I0)}, {\bf (I1a)} and {\bf (I2)}. This is due to the fact that we
are showing a strict Mourre estimate, i.e.\ without compact
contribution.

We now give the plan of the paper. In Section \ref{s:Pauli}, we recall
some definitions and properties of Pauli-Fierz models. In Section
\ref{s:conjop}, we construct the conjugate operators. In Section
\ref{s:regu}, we prove the regularity properties so that one may
apply the Mourre theory. The Virial theorem is discussed in Section
\ref{s:virial}. In Section \ref{s:far}, we establish the extended
Mourre estimate far from the thresholds for small coupling
constants, we explain in Remark \ref{r:motiv} why the method should be
improved to obtain the result above a threshold. In Section
\ref{s:thre},  we settle the extended Mourre estimate above the
thresholds under the hypotheses of a Fermi golden rule. In Appendix
\ref{s:level}, we explain how to check the Fermi golden rule and why
this hypothesis is compatible with the hypothesis {\bf (I0)},
{\bf (I1a)}, {\bf   (I1b)} and {\bf (I2)}. In Appendix
\ref{s:semiprop}, we gather some properties of $C_0$-semigroups and in
Appendix \ref{s:ms} we  recall the properties of the $\Cc^1$ class in
this setting and the hypotheses so as  to apply the extended
Mourre theory.

\subsection*{Notation:} Given a borelian set $\Jc$, we denote by
$E_\Jc(A)$ the spectral measure associated to a self-adjoint
operator $A$ at energy $\Jc$. Given  Hilbert spaces $\Hr, \Kr$, we
denote by $\Bc(\Hr, \Kr)$ the set of bounded operator from $\Hr$ to
$\Kr$. We simply write $\Bc(\Hr)$, when $\Hr=\Kr$. We denote by
$\sigma(H)$ the spectrum of $H$. We set $\langle x
\rangle:=(1+x^2)^{1/2}$.  We denote by $\|\cdot\|_\Hr$  and by
$\langle \cdot, \cdot \rangle_\Hr$ the norm and the scalar product of
$\Hr$, respectively. We omit the indices when no confusion 
arises. We denote by $\wlim$ and $\slim$ the weak and strong limit,
respectively. A dot over a Besov or a Sobolev space denotes the closure
of the set $\Cc_c^\infty$  of smooth functions with compact support,
with respect  to the norm of the space.

\subsection*{Acknowledgments:} I express my gratitude 
to Jan Derezi\'nski who encouraged me in pursuing these ideas. I would
also like to thank Volker Bach, Alain Joye, Christian G\'erard,
Vladimir Georgescu, Wolfgang Spitzer, Claude-Alain Pillet and Zied
Ammari for some useful discussions. This work was partially supported
by the Postdoctoral Training Program HPRN-CT-2002-0277.  

\section{The Pauli-Fierz model}\label{s:Pauli}
\setcounter{equation}{0}
Pauli-Fierz operators are often used in quantum physics as generator
of approximate dynamics of a (small) quantum system interacting with a free
Bose gas. They describe typically a non-relativistic atom interacting
with a field of massless scalar bosons. The quantum system is given by a (separable) complex Hilbert space
$\Kr$. The Hamiltonian describing the system is denoted by a
self-adjoint operator $K$, which is bounded from below. We will
suppose that $K$ has some discrete spectrum. One may consider purely
discrete spectrum, like \cite{GGM2}, or not, like in \cite{Ski}. 
To do not mutter the presentation, we will 
take $\Kr= {\rm Ran\, } E_{\Ic}(K)$, where $\Ic$ contains a finite number of 
eigenvalues and consider the restriction of $K$ to this space. 
Hence, we restrict the analysis to a self-adjoint matrix $K$
acting in a Hilbert space $\Kr$ of finite dimension. This corresponds
to analyze  $n$ level atoms. Doing so, we avoid some light problems of
domains, which are already discussed in details in \cite{GGM2, Ski}
and gain in clarity of presentation.   

\subsection{The bosonic field}\label{s:bosonic}
We refer to \cite{BR, BSZ, RS} for a more thorough discussion of these
matters.
The bosonic field is described by the Hilbert space $\Gamma(\hg)$, where
$\hg$ is a Hilbert space. We recall its construction. Set
$\hg^{0\otimes}=\C$ and $\hg^{n\otimes}=\hg\otimes \dots \otimes
\hg$. Given a closed operator $A$, we define the closed operator
$A^{n\otimes}$ defined on $\hg^{n\otimes}$ by $A^{0\otimes}=1$ if
$n=0$ and by $A\otimes \ldots \otimes A$ otherwise. Let $S_n$ be the
group of permutation of $n$ elements. For each $\sigma\in S_n$, one
defines the action on $\hg^{n\otimes}$ by
$\sigma(f_{i_1}\otimes\ldots\otimes f_{i_n})= f_{\sigma^{-1}(i_1)}\otimes
\ldots\otimes f_{\sigma^{-1}(i_n)}$, where $(f_i)$ is a basis of $\hg$.  
The action extends to $\hg^{n\otimes}$ by linearity to a unitary
operator. The definition is independent of the choice of the
basis. On $\hg^{n\otimes}$, we set
\begin{eqnarray}\label{e:Pn}
\Pi_n:=\frac{1}{n!}\sum_{\sigma\in S_n} \sigma \mbox{ and }
\Gamma_n(\hg):= \Pi_n(\hg^{n\otimes}).
\end{eqnarray} 
Note that $\Pi_n$ is an orthogonal projection. We call $\Gamma_n(\hg)$
the $n$-particle bosonic space. The bosonic space is defined
by
\begin{eqnarray*}
\Gamma(\hg):=\bigoplus_{n=0}^{\infty} \Gamma_n(\hg).
\end{eqnarray*} 
We denote by $\Omega$ the \emph{vacuum}, the element
$(1,0,0,\ldots)$ and by $P_\Omega:=\Gamma(\hg)\rightarrow
\Gamma_0(\hg)$ the projection  
associated to it. We define $\Gamma_{\rm fin}(\hg)$ the set of finite
particle vectors, i.e.\ $\Psi=(\Psi_1, \Psi_2, \ldots)$ such that
$\Psi_n=0$ for $n$ big enough.

We now define the second quantized operators. We recall that a densely
defined operator $A$ is closable if and only if its adjoint $A^*$ is
densely defined. Given a closable operator $q$ in $\hg$. We define
$\Gamma_{\rm fin}(q)$ acting from $\Gamma_{\rm fin}(\Dc(q))$ into
$\Gamma_{\rm fin}(\hg)$ by  
\begin{eqnarray*}
\Gamma_{\rm fin}(q)|_{\Pi_n( \Dc(q)^{n\otimes})}:= q\otimes \ldots
\otimes q.  
\end{eqnarray*}  
Since $q$ is closable, $q^*$ is densely defined. Using that
$\Gamma_{\rm fin}(q^*)\subset \Gamma_{\rm fin}(q)^*$, we see that
$\Gamma_{\rm fin}(q)$ is closable and we denote by $\Gamma(q)$ its
closure. Note  that $\Gamma(q)$ is bounded if and only if $\|q\|\leq
1$.  

Let $b$ be a closable operator on $\hg$. We define $\dG_{\rm fin}(b):\Gamma_{\rm fin}\big(\Dc(b)\big)\longrightarrow
\Gamma_{\rm fin}(\hg)$ by
\begin{eqnarray*}
\dG_{\rm fin}(b)|_{\Pi_n( \Dc(b)^{n\otimes})}:= \sum_{j=1}^{n} 1\otimes
\ldots \otimes 1 \otimes \underbrace{b}_{j^{\rm th}} \otimes  1\otimes
\ldots \otimes 1.
\end{eqnarray*} 
As above, $\dG_{\rm fin}(b)$ is closable and $\dG(b)$ denotes also
its closure. We link the objects.
\begin{lemma}\label{l:semi}
Let $\R^+\ni t\mapsto w_t\in\Bc(\hg)$ be a $C_0$-semigroup of
contractions (resp.\ of isometries), with generator $a$. Then $\R^+\ni
t\mapsto \Gamma(w_t)\in\Bc\big(\Gamma(\hg)\big)$ is a $C_0$-semigroup of
contractions (resp.\ of isometries)  whose generator is $\dG(a)$.
\end{lemma} 
\proof It is easy to see that $W_t:=\Gamma(w_t)$ is a $C_0$-semigroup
of contractions (resp.\ of isometries). Let $A$ be its
generator. Immediately, one gets $\dG_{\rm fin}(a)\subset A$.  Since
$\Gamma_{\rm fin}\big(\Dc(a)\big)$ is dense in $\dG(\hg)$ and invariant under
$W_{t}$, the Nelson lemma gives that $\Gamma_{\rm fin}\big(\Dc(a)\big)$ is
dense in $\Dc(A)$ for the graph norm and also that $\dG(a)=A$.\qed  

\subsection{The interacting system}\label{s:inter} Given a self-adjoint operator
$\omega$ in $\hg$ and a finite dimensional Hilbert space $\Kr$. 
One defines the free Hamiltonian $H_0$ acting on the Hilbert space
$\Hr:=\Kr\!\otimes \Gamma(\hg)$ by  
\begin{eqnarray}\label{e:H0}
H_0:=K\otimes 1_{\Gamma(\hg)}+1_\Kr\! \otimes \dG(\omega).
\end{eqnarray} 
We recall also the definition of the \emph{number
operator} $N:=1_\Kr\!\otimes\dG(\id)$.  

We now define the interaction. Let $\alpha$ be an element $\Bc(\Kr, \Kr\!\otimes
\hg)$. This is a \emph{form-factor}. We define $b(\alpha)$ on $\Hr$ by
$b(\alpha):= \Kr\!\otimes h^{n\otimes}\rightarrow \Kr\!\otimes
h^{(n-1)\otimes}$, where  
\begin{eqnarray*}
b(\alpha)(\Psi\otimes \phi_1\otimes \ldots \phi_n):=\alpha^*(\Psi\otimes
\phi_1)\otimes \phi_2\otimes \ldots \phi_n,
\end{eqnarray*}  
for $n\geq 1$ and by $0$ otherwise. This operator is bounded and its 
norm is given by $\|\alpha\|_{\Bc(\Kr, \Kr\!\otimes \hg)}$. 
We define the \emph{annihilation operator} on $\Kr\!\otimes \Gamma(\hg)$
with domain $\Kr\!\otimes \Gamma_{\rm fin}(\hg)$ by   
\begin{eqnarray*}
a(\alpha):=(N+1)^{1/2}b(\alpha)\big(1\otimes \Pi), 
\end{eqnarray*} 
where $\Pi:=\sum_n \Pi_n$, see \eqref{e:Pn}. As above, it is
closable and its closure is denoted by $a(\alpha)$. Its adjoint is the
\emph{creation operator}. It acts as $a^*(\alpha)= b^*(\alpha)
(N+1)^{1/2}$ on $\Hr$. Note that $b^*(\alpha)(\psi\otimes\phi_1\otimes \ldots
\otimes \phi_n)=(\alpha\phi)\otimes \phi_1\otimes \ldots  \otimes
\phi_n$. 

The \emph{(Segal) Field operator} is defined by 
\begin{eqnarray*}
\phi(\alpha):=\frac{1}{\sqrt{2}}\big(a(\alpha)+a^*(\alpha)\big).
\end{eqnarray*} 
We consider its closure on $\Kr\!\otimes \Dc(N^{1/2})$. We have
the two elementary estimates:
\begin{eqnarray}\label{e:bdN}
\|(N+1)^{-1/2}a^{(*)}(\alpha)\|\leq \|\alpha\|, \quad
\|(N+1)^{-1/2}\phi(\alpha)\|\leq \sqrt{2}\|\alpha\|. 
\end{eqnarray} 
An assertion containing $(*)$ holds with and without $*$.

We give the following $N_\tau$-estimate and refer to \cite[Proposition
4.1]{DJ} for a proof of i). The point ii) is a direct consequence of
the Kato-Rellich Lemma. This kind of estimates comes back to \cite{GJ}. 
See also \cite{BFS}. We refer to \cite{G}[Appendix A] and
\cite{GGM2}[Proposition 3.7] for unbounded $K$.  

\begin{proposition}\label{p:ntau}
Let $\omega$ be a non-negative, injective, self-adjoint operator on
$\hg$. Let $\beta\in\Bc\big(\Kr, \Kr\!\otimes \Dc(\omega^{-1/2})\big)$. 

{\rm i)} Then
$\phi(\beta)\in\Bc\big(\Kr\!\otimes\Dc(\dG(\omega)^{1/2}), \Hr\big)$ 
and for any $\Phi\in \Dc(\dG(\omega)^{1/2})$, 
\begin{align}
\nonumber
|\phi(\beta)\Phi\|^2\leq& \|\beta\|_{\Bc(\Kr, \Kr\!\otimes \hg)} \,
\|\Phi\|^2
\\\label{e:ntau0} & + 2\|\omega^{-1/2}\beta\|_{\Bc(\Kr, \Kr\!\otimes \hg)} \,
\langle \Phi,  \un_\Kr\!\otimes\dG(\omega) 
\Phi  \rangle.   
\end{align}  

{\rm ii)} The field operator $\phi(\alpha)$ is $H_0$-operator bounded
with relative bound $\varepsilon $, for all $\varepsilon >0$. Hence,
$H_\lambda:= H_0 +\lambda \phi(\alpha)$, for $\lambda\in\R$,
defines a self-adjoint operator with domain
$\Dc(H_\lambda)=\Kr\!\otimes\dG(\omega)$ and 
is essentially self-adjoint on any core of $H_0$.
\end{proposition}

\subsection{The zero-temperature Pauli-Fierz Model}\label{s:model} We
now precise our model to the zero-temperature physical setting. The
one particle space is given by $\hg:=L^2(\R^d, dk)$, where $k$ is the
boson momentum. The \emph{one particle kinetic energy} is the operator
of multiplication by  $\omega(k):=|k|$. Consider a self-adjoint matrix
$K$ on a finite dimensional Hilbert space $\Kr$ and denote by
$(k_i)_{i=0, \ldots, n}$, with $k_i\pp k_{i+1}$ its eigenvalues. We
denote by $P_{k_i}$ the projection onto the $i$-th eigenspace.  

The spectrum of $\dG(\omega)$ in $\Gamma(\hg)$ is $[0, \infty)$ and
due the vacuum part, $0$ is the only eigenvalue. Its multiplicity is one.
The spectrum of $H_0$ given by \eqref{e:H0} is $[k_0,
\infty)$. The eigenvalues are   given by  $(k_i)_{i=0, \ldots, n}$
and have the same multiplicity as those of $\Kr$. The singularly 
continuous component of the spectrum is  empty. Here, $(k_i)_{i=0,
  \ldots, n}$ play also the r\^ole of \emph{thresholds}. 

We consider a form-factor $\alpha$ satisfying hypothesis {\bf
  (I0)}. By applying Proposition \ref{p:ntau}, the operator  
$H_\lambda$, given by \eqref{e:H}, is self-adjoint and
$\Dc(H_{\lambda})= \Kr\!\otimes \Dc\big(\dG(\omega)\big)$. 

Since we study form factors in $\Bc(\Kr,\Kr\!\otimes \hg)$, we
forbid some eventual singularities of the form-factor from the very
beginning. However, if the atomic part has a particular shape, one may
use some gauge transformations and gains in singularity, see for
instance \cite{GGM2}[Section 2.4] and \cite{DJ}[Section 1.6]. Nevertheless,
it is an open question if there exists some gauge
transformation that allows one to cover the physical form factor studied
in \cite{BFS, BFSS}, from our conditions.  Conversely, the
classes of perturbations studied in the latter does not fully cover ours.  

\section{The conjugate operators}\label{s:conjop}
\setcounter{equation}{0}
In this paper, we analyze the spectrum of the Pauli-Fierz Hamiltonian
$H_\lambda$ described in Section \ref{s:model} using some commutator
techniques. We study the behavior of the embedded  eigenvalues of
$H_\lambda$ under small coupling constants and  establish some refined
spectral properties. To do so, we establish a version of the Mourre
estimate, see Appendix \ref{s:LAPhyp}. Hence, we start by constructing
the conjugate operator. We follow similar ideas as 
in \cite{GGM2, hsp, Ski}. Later, we modify it by a finite rank
perturbation, in the spirit of \cite{BFSS}. Unlike in the standard
Mourre theory, the conjugate operator is not self-adjoint and only
maximal-symmetric. We refer to Appendix \ref{s:C1} for discussions
about $1$-commutators properties in this setting. We point out that 
one may avoid to work with maximal-symmetric operator by symmetrizing
the space and thus gluing non-physical free bosons, see
\cite{DJ}[Section 5.2].  This trick leads to some   
problems of domains with our method and would be treated elsewhere.  

We point out that the real drawback of this choice of conjugate
operator comes from the fact that the commutator is not $H_\lambda$
bounded, like in the standard Mourre theory and \cite{BFS, BFSS, FGS, FP}. Some
difficulties appear to apply the Virial theorem. To overcome them, we follow ideas 
of \cite{Ski, GGM2} and construct a series
of approximate conjugate operators.  One may also proceed like in \cite{M}.    

\subsection{The semigroup on the $1$-particle space}
Fix $\cchi\in\Cc^\infty_c\big(\R^+;[0,1]\big)$ decreasing such that
$\cchi(x)=1$ for $x\leq 1$ and $0$ for $x\geq 2$. Set $\tilde \cchi :=
1-\cchi$. We consider the following vector fields on $\R^+$:
\begin{align}\label{e:sd}
m_n(t):=
\begin{cases}
\tilde \cchi\left(n t\right), & \mbox{ for } n\in\N,
\\
1, & \mbox{ for } n=\infty, 
\end{cases} \mbox{ and } s_n(t)=\frac{m_n(t)}{t}. 
\end{align}
Note that $m_n$ converges increasingly to $m_\infty$, almost
everywhere, as $n$ goes to infinity.  As in \cite{Ski} and in
\cite{GGM2}, the r\^ole of $m_\infty$ would be to ensure the positivity of
the commutator and the one of $m_n$ would be to guarantee of the
Virial theorem.  

We define the associated vector fields in $\R^d$ as follows:
\begin{eqnarray}\label{e:sdv}
\overrightarrow{s_n}(k) :={s}_n \big(|k|\big)k, \mbox{ for } k\in\R^d
\mbox{ and } 
n\in\N^*\cup\{\infty\}.
\end{eqnarray}
We shall construct the $C_0$-semi\-group of isometries associated to
the vector fields $\overrightarrow{s_n}$ on $\hg=L^2(\R^d)$ and
identify the generators. We define  
\begin{eqnarray}\label{e:sgp}
a_n:=\displaystyle-\frac{1}{2}\Big(\overrightarrow{s_n}\cdot D_k + 
D_k \cdot\overrightarrow{s_n} \Big)
\end{eqnarray} 
on $\Cc^\infty_c(\R^d\setminus \{0\})$ for all
$n\in\N^*\cup\{\infty\}$ and where $D_k=i\nabla$. These operators are
closable as the domains of their adjoints are dense. In the sequel, we
denote by the same symbol their closure. 

We work in polar coordinates. We identify $\hg$ and $\tilde \hg$ through the
transformation \eqref{e:T}. Given an operator $B$ in $\hg$, we denote by
$\tilde B$ the corresponding operator acting in the $\tilde\hg$ and
given by $\tilde B:= T B T^{-1}$. We have:  
\begin{proposition}\label{p:ess}
For $n$ finite, $a_n$ is essentially self-adjoint on
$\Cc^\infty_c(\R^d\setminus \{0\})$ and $a_\infty$ is maximal
symmetric with deficiency indices $(N, 0)$. Here, $N=\infty$ for
$d\geq 2$ and $N=2$ for $d=1$. The operator $a_n$ generates a
$C_0$-semigroup of isometries denoted by
$\{w_{n,t}\}_{t\in\R^+}$. In polar coordinates, the domains are given by 
\begin{align*}
\Dc(\tilde a_n)\supset \Dc(\tilde a_\infty)=&\, \dot\Hr^1(\R^+)\otimes
L^2(S^{d-1}), \mbox{ for all } n\in \N^*,
\\
\Dc(\tilde a_\infty^*)=&\, \Hr^1(\R^+)\otimes L^2(S^{d-1}),
\end{align*} 
where $\dot\Hr^1(\R^+)$ is the closure of $\Cc^\infty_c(\R^{+})$ under
the norm $\|\cdot \|+\|\partial_r \cdot\|$ and where $\Hr^1(\R^+)$
is the  Sobolev space of first order. 
\end{proposition}
See Section \ref{s:semiprop} for an overview on $C_0$-semigroups. For $n$ finite, the $C_0$-semigroup extends to a $C_0$-group since $a_n$ is self-adjoint. 

\proof When $n$ is finite, it is well known that  $a_n$ is essentially
self-adjoint on $\Cc^\infty_c(\R^d)$ and follows by studying $C_0$-group associated to the  
flow defined by the smooth vector field $\overrightarrow{s_n}$. The
density follows by the Nelson lemma. See for instance
\cite{ABG}[Proposition 4.2.3]. Hence, for $n$ finite, it remains  to show
that $\Cc^\infty_c(\R^d\setminus \{0\})$ is a core for $a_n$. 

Straightforwardly, for $n\in\N^*\cup\{\infty\}$, one gets
\begin{eqnarray}\label{e:atilde}
\quad \quad {\tilde a}_n:=Ta_n T^{-1}= i\big(m_n(\cdot)\partial_r
+\frac{1}{2} (m_n)'(\cdot) \big)\otimes \un, \mbox{ where } m_n(r):=r
s_n(r). 
\end{eqnarray}

We extend $m_n$ on $\R$ by setting $m_n(-r):=m_n(r)$
for $r\pg 0$ and prolongate it by continuity in $0$. Let $\phi_{n,t}$ be
the flow generated by the smooth vector field $m_n$ on $\R$. In other
words, $\phi_{n,t}:=\phi_n(t,\cdot)$ is the unique solution of 
$(\partial_t\phi_n)(t,r)= m_n\big(\phi_n(t,r)\big)$, where  $\phi_n(0,r)=r$.
Since $m_n$ is globally Lipschitz, $\phi_{n,t}$ exists for all time $t$.  
Moreover, $\phi_{n,t}$ is a smooth diffeomorphism of $\R$ with inverse
$\phi_{n,-t}$ for all $t\in\R$. Let $\tilde \phi_{n,t}$ be the restriction
of $\phi_{n,t}$ from $\R^{+*}$ to $\R^{+*}$. Let  $\Omega_{n,t}$ be the domain
of this restriction, i.e.\ the set of $r\pg 0$ such that $\phi_{n,t}(r)\pg
0$. One has $\Omega_{n,t}=\R^{+*}$ for $t\geq 0$ as $m_n(r)$ is
positive. For the same reason, $t\mapsto \Omega_{n,t}$ is increasing. Note
also that we have $\Omega_{n,-t}=\phi_{n,t}(\R^{+*})$ for $t\geq 0$. 
For $u\in\tilde \hg$, we  set:
\begin{eqnarray}\label{e:wtilde}
(\tilde w_{n,t} u)(r, \theta) := 1_{\Omega_{n,-t}}(r)\sqrt{\phi_{n,-t}'(r)}
u(\phi_{n,-t}(r),  \theta), \mbox{ for } t\geq 0.  
\end{eqnarray}
A change of variable gives that $\tilde w_{n,t}$ is an isometry of
$L^2(\R^+)$ with range $L^2(\Omega_{n,-t})$ for all $t\geq 0$. Since
$\phi_{n,t}$ is a smooth flow, $\{\tilde w_{n,t}\}_{t\geq 0}$ is a
$C_0$-semigroup of isometries. The adjoint $C_0$-semigroup is given by 
\begin{eqnarray}\label{e:wadj}
(\tilde w_{n,t}^*u)(r, \theta) := 1_{\R^{+*}}(r)	\sqrt{\phi_{n,t}'(r)}
u(\phi_{n,t}(r), \theta), \mbox{ for } t\geq 0.  
\end{eqnarray}
This is \emph{not} a semigroup of isometries when $n=\infty$. 

We compute the generator of the semigroup $\{\tilde w_{n,t}\}_{t\geq
0}$. Take $u\in \Cc^\infty_c\big(\tilde \hg\big)$. We have $\tilde
w_{n,t} u\in \Cc^\infty_c(\Omega_{n,-t}\times S^{d-1})$. Let $r\in
\Omega_{n,-t}$, we get 
\begin{eqnarray*}
-\left(\frac{d}{dt} \tilde w_{n,t} u\right)(r, \theta)=	\left(\tilde w_{n,t}
 \big(m_n(\cdot) \partial_r +\frac{1}{2} (m_n)'(\cdot)\big)u
 \right)(r,\theta). 
\end{eqnarray*} 
Denoting by the same symbol the closure of 
$\tilde a_n$ on $\Cc^\infty_c(\R^{+*}\times S^{d-1})$, we obtain
\begin{eqnarray*}
-i \frac{d}{dt} \tilde w_{n,t} u= \tilde w_{n,t} \tilde a_n u.
\end{eqnarray*} 
The closed operator is \emph{a priori} only a restriction of the
generator of the semigroup (in the sense of the inclusion of graph of
operators). Now, since $\tilde w_{n,t}$ stabilizes
$\Cc^\infty_c(\tilde \hg)$ for all $t\geq 0$, the Nelson
lemma gives that this space is a core for generator of the
$C_0$-semigroup $\{\tilde w_{n,t}\}_{t\geq 0}$. Since this one is an
extension of $\tilde a_n$, we have shown that $\tilde a_n$ is really
the generator. One may denote formally $\tilde w_{n,t}=e^{it\tilde
a_n}$. The domain of $\tilde a_n$ contains $\dot\Hr^1(\R^+)\otimes
L^2(S^{d-1})$. Easily, this is an equality for $n=\infty$. 

Considering the spectrum of $a_n$, we derive the deficiency
indices of the closure of $a_n$  on $\Cc^\infty_c(\R^d\setminus\{0\})$
are of the form $(N,0)$. For $n$ finite these indices are equal, we
infer the essential self-adjointness of  $a_n$ on
$\Cc^\infty_c(\R^d\setminus\{0\})$. 

At this point, one may feel the real difference between the case $n$
finite and infinite. On one hand $m_\infty\geq 1$ and on the other
hand, for finite $n$, $m_n(r)$ tends to $0$ as $r$ tends to $0$. The
domain of the adjoint of $\tilde a_\infty$ would be different. Indeed, 
\begin{eqnarray}\label{e:aadj}
\big(\tilde a_\infty^*u\big)(r, \theta)= i\big(m_\infty(r)(\partial_r
u)(r,\theta) +\frac{1}{2} (m_\infty)'(r)u(r,\theta) \big), 
\end{eqnarray}
where $u\in\Dc(\tilde a_\infty^*)=\Hr^{1}(\R^+)\otimes
L^2(S^{d-1})$. Moreover, when $n=\infty$, the deficiency indices are then
$(\infty, 0)$, as the dimension of $L^2(S^{d-1})$ is infinite. \qed

\subsection{The $C_0$-semigroup on the Fock space.}

Thanks to Proposition \ref{p:ess} and Lemma \ref{l:semi}, we define
the $C_0$-semigroups on the whole Hilbert space. We set:  
\begin{eqnarray}\label{e:semi}
W_{n,t}:=\un_\Kr\!\otimes \Gamma(w_{n,t}) \mbox{ and } W^{*}_{n,t}=\un_\Kr\!\otimes
\Gamma(w^{*}_t), \mbox{ for } t\geq 0. 
\end{eqnarray}
Clearly, $\{W_{n,t}\}_{t\geq 0}$ is a $C_0$-semigroup of isometries. Let
$A_\infty$ be its generator.  In the same way, for $n$ finite, we set  
\begin{eqnarray}\label{e:semin}
A_n:=\un_\Kr\!\otimes \dG(a_n).
\end{eqnarray}
This is the generator of the $C_0$-group $\un_\Kr\!\otimes\Gamma
(e^{ita_n})$ by Lemma \ref{l:semi}. Recall the r\^ole of the $A_n$ is to
ensure a Virial theorem, see Proposition \ref{p:6.5}. 

In Section \ref{s:far}, we see that the operator $A_\infty$ alone is
not enough to deal with threshold energy as the system could be
uncoupled. One needs to take in account the Fermi golden rule. One way
is to follow \cite{DJ} and to take advantage of it in the limiting
absorption principle. Another way is to modify the conjugate operator
with a \emph{finite rank   perturbation} so as to obtain more
positivity above the thresholds, by letting appearing the Fermi golden
rule in the commutator, see Section \ref{s:thre}. This idea comes from
\cite{BFSS}. We follow it. 

Choose  $k_{i_0}$ an eigenvalue of $K$ and assume that \eqref{e:FGR0}
holds true at energy $k_{i_0}$ for the couple  $(H_0, \alpha)$. 
Let $P$ be the projector  $P_{k_{i_0}}\otimes P_\Omega$. For
$\varepsilon<\varepsilon_0$, we define   
\begin{eqnarray*}
\hat A_n:=A_n+\lambda \theta B_\varepsilon, \mbox{ for }
n\in \N^*\cup\{\infty\}, 
\end{eqnarray*} 
where $B_\varepsilon :=\im (\overline{R_\varepsilon}^2\phi(\alpha)P)$,
$R_\varepsilon:=\big((H_0-k_{i_0})^2+\varepsilon^2\big)^{-1/2}$ and 
$\overline{R_\varepsilon}:=\overline{P} R_\varepsilon$.
Note that the conjugate operator depends on the two parameters
$\lambda\in \R$ from the coupling constant, $\varepsilon>0$ from the
Fermi golden rule hypothesis and on an extra technical $\theta>0$.  
For the sake of clarity, we do not write these extra dependences. 

Using Proposition \ref{p:kato} and the fact that $B_\varepsilon $ is 
bounded, one gets $\hat A_\infty$ is the generator of
a $C_0$-semigroup. A bit more is true.  

\begin{lemma}\label{l:semiA}
The operator $\hat A_\infty$ is maximal symmetric on
$\Dc(A_\infty)$ and is the generator of $C_0$-semigroup of
isometries, denoted by  $\{\widehat W_{n,t}\}_{t\geq 0}$. For $n$
finite, the operator $\hat A_n$ is self-adjoint on the domain of
$\Dc(A_n)$.  
\end{lemma}

\proof
The second point is obvious. We concentrate on the first one. By
Proposition \ref{p:ess}, $A_\infty$ is maximal symmetric with deficiency
indices $(N,0)$ for some $N\neq 0$. Since $B_\varepsilon$ is bounded,
there is $c<0$ such that $\|B_\varepsilon(A_\infty -z)^{-1}\|<1$, for all
$z\in \C$ where $\im(z)\leq c$. Since $(I+ B_\varepsilon(A_\infty
-z)^{-1})(A_\infty -z)= A_\infty+B_\varepsilon-z$ on the domain of
$A_\infty$,  we  get the spectrum of  $\hat A_\infty$ is contained in an
upper half plane $\R + i [c, \infty)$. Now, since  $B_\varepsilon$ is
symmetric, so is $\hat A_\infty$. If the indices of $\hat A_\infty$
would be both non-zero then its spectrum would be $\C$. Therefore, the
deficiency indices of $\hat A_\infty$ are $(N',0)$ for 
some non-negative $N'$. Note that $N'\neq 0$ by the
Kato-Rellich theorem applied on $\hat A_\infty$, since $B_\varepsilon$ is
bounded. Hence, $\hat A_\infty$ is maximal symmetric on $\Dc(A_\infty)$ and its
spectrum is $\R + i [0, \infty)$. It is  automatically a
$C_0$-semigroup of isometries. \qed

\section{Smoothness with respect to the $C_0$-semigroup}\label{s:regu}
\setcounter{equation}{0}
In Section \ref{s:gene}, we recall a general result. In
Section \ref{s:1comm}, we give some $1$-commutator properties for
$A_n$. We check the hypothesis  {\bf (M1)}--{\bf (M4)} of Appendix
\ref{s:LAPhyp}. We identify the spaces and operators appearing therein
in Lemma \ref{l:6.4}. In Section \ref{s:1comm'}, we extend these
properties to $\hat A_n$, using Proposition \ref{p:kato} and
Lemma \ref{l:finite}. The Virial theorem is discussed in Section
\ref{s:virial}.  At last, second commutator assumptions and the
hypothesis {\bf (M5)} are discussed in Section \ref{s:2comm}. 

 \subsection{A general result}\label{s:gene} In order to check the
$C^1$ properties, the $b$-stability, see Definition \ref{d:bstable},
and to be able to deduce hypothesis {\bf (M1)-(M5)} of Appendix
\ref{s:LAPhyp}, we recall \cite{GGM2}[Proposition 4.10]. We formulate
it for bounded $K$. Set first a $C_0$-semigroup of isometries $\R^+\ni
t \rightarrow v_t\in\Bc(\hg)$ with generator  $a$. By Lemma
\ref{l:semi}, $V_t:=\un_\Kr\!\otimes \Gamma(v_t)$ is a $C_0$-semigroup
of isometries with generator $A=\un_\Kr\! \otimes\dG(a)$. Let $b\geq 0$
be a self-adjoint operator on $\hg$, and $K$ as in \eqref{e:H0}. Set      
\begin{eqnarray*}
 B:=K\otimes \un_{\Gamma(\hg)} + \un_\Kr\! \otimes \dG(b), \quad
 \Gr_B:=\Dc(B^{1/2})=\un_\Kr\! \otimes \Dc\big(\dG(b)^{1/2}\big).  
\end{eqnarray*} 
\begin{proposition}\label{p:4.10} Let $\omega$ and $b\geq 0$ acting in
$\hg$. Then,

{\rm i)} The space $\Gr_B$ is $b$-stable under  $\{V_t\}_{t\in\R^+}$
  {\rm (}resp.\  $\{V_t^*\}_{t\in\R^+}${\rm )}, if
\begin{eqnarray}\label{e:4.19}
v^*_t b v_t\leq C_t b, (\mbox{resp.\ } v_t bv^*_t \leq C_t b) \mbox{
with } \sup_{0<t<1}C_t<\infty.
\end{eqnarray} 

{\rm ii)}  Assuming \eqref{e:4.19} and that there is a constant $C$ such that  for all $u_i\in\Dc(b^{1/2})$
\begin{eqnarray}\label{e:4.21}
\quad \quad \omega\leq Cb,\,\, |\langle u_2, (\omega v_t-v_t\omega) u_1\rangle |\leq 
Ct\|b^{1/2}u_1 \|\cdot \|b^{1/2}u_2\|, \mbox{ for } 0<t<1.
\end{eqnarray}
Then $H_0\in\Cc^1(A; \Gr_B, \Gr^*_B)$. Besides, in the sense of forms on $\Gr_B$, one has
\begin{eqnarray*}
[H_0, iA]^\circ=\un_\Kr\!\otimes \dG([\omega, ia]^\circ).
\end{eqnarray*} 

{\rm iii)} Assume \eqref{e:4.19} and that $\alpha$ is a
form-factor satisfying 
\begin{eqnarray}\label{e:4.22}
\alpha\in\Bc\big(\Kr, \Kr\!\otimes \Dc(a)\big),\, \,  a\alpha \in
\Bc\big(\Kr, \Kr\!\otimes \Dc(b^{-1/2})\big).
\end{eqnarray} 
Then $\phi(\alpha)\in\Cc^1(A; \Gr_B, \Gr_B^*)$ and in the sense of
forms on $\Gr_B$, we get 
\begin{eqnarray*}
[\phi(\alpha), iA]^\circ= -\phi(ia\alpha).
\end{eqnarray*} 
\end{proposition} 
Here $[\cdot, \cdot]^\circ$ denotes the closure of the form defined by
$[\cdot, \cdot]$, $H_0$ is defined in \eqref{e:H0} and $a\alpha$ is a
short for $(1\otimes a) \alpha$. If {\bf (I0)} and
\eqref{e:4.19}--\eqref{e:4.22} hold true, 
then $H_\lambda$, defined in \eqref{e:H}, is self-adjoint with the
same domain as $H_0$ and lies in $\Cc^1(A;  \Gr_B, \Gr^*_B)$.   

\subsection{Estimation on the first commutator}\label{s:1comm}
In this section, we compute the first commutator with respect to the
conjugate operator $A_\infty$ and check the hypotheses {\bf (M1)}-{\bf
  (M4)} discussed in Appendix \ref{s:LAPhyp}. We follow \cite{GGM}
and use only the hypotheses {\bf(I0)} and {\bf(I1a)}. We start with a
direct consequence of  Proposition \ref{p:ess}.  
\begin{lemma}\label{l:6.2}
We assume {\bf (I0)} and {\bf (I1a)}. Then, $\alpha\in\Bc\big(\Kr, \Kr\!\otimes
	\Dc(a_n)\big)$ and $a_n \alpha\in \Bc\big(\Kr, \Kr\!\otimes
	\Dc(\omega^{-1/2})\big)$, for all $n\in \N^*\cup\{\infty\}$.
\end{lemma}
We formally decompose the commutator $[H_\lambda, iA_n]$ into two parts. We set:
\begin{eqnarray}\label{e:M&R}
\begin{cases}
 M_n:=\un_\Kr\!\otimes \dG(m_n) + \un \otimes P_\Omega,
\\
S_n:=-\phi(ia_n \alpha) - \un_\Kr\! \otimes P_\Omega, 
\end{cases}\quad  \mbox{ for all } n\in \N^*\cup\{\infty\}.
\end{eqnarray}
Here, we add $\un_\Kr\!\otimes P_\Omega$ to obtain $M_\infty\geq
1$. We stress that, for finite $n$, $M_n$ has a different domain as
 $M_\infty$. Indeed,  $\Dc(M_n)\subset \Dc(H_0)$ when $n$ is
finite and  $\Dc(M_\infty) = \Dc(N)$, since
$M_\infty=N + \un_\Kr\!\otimes P_\Omega$. 

We start with the hypothesis {\bf (M1)}. We need to precise the
definition the commutator $H'_\lambda$  given formally by $[H_\lambda,
iA_\infty]$. Note that it does not extend to a $H_\lambda$-bounded
operator, as in the standard Mourre theory. We follow \cite{GGM} and define
\begin{eqnarray}\label{e:B}
B_\infty:= K\otimes \un_{\Gamma(\hg)}+\un_\Kr\!\otimes
\dG\big((k^2+1)^{1/2}\big).
\end{eqnarray} 
Let $\Dr_\infty:=\Dc(B_\infty)$ and
$\Gr_\infty:=\Dc(B^{1/2}_\infty)$. We would drop the subscripts after
this lemma as no more confusion could arise with Appendix \ref{s:LAPhyp}. 

\begin{lemma}\label{l:6.4}
Assume {\bf (I0)} and {\bf (I1a)}. Then:

{\rm i)} $H_\lambda\in\Cc^1(M_\infty)$, $\Dc(H_\lambda)\cap
\Dc(M_\infty)$ is a core for $M_\infty$, $S_\infty$ is  symmetric and
lies in $\Bc(\Dc(H_0), \Hr)$.  

{\rm ii)} Let $H'_\lambda$ be the closure of 	$M_\infty+S_\infty$
  defined on   	$\Dc(H_\lambda)\cap \Dc(M_\infty)$. Therefore, $H_\lambda$ and 
$H'_\lambda$ satisfy {\bf (M1)}. 

{\rm iii)} $\Dr_\infty=\Dc(H'_\lambda)\cap \Dc(H_\lambda)
=\Dc(M_\infty)\cap \Dc(H_\lambda)$ and $\Gr_\infty$ is the same as in \eqref{e:G}.  
\end{lemma}
\proof We start with i). Take the $C_0$-group generated  by $m_\infty$
acting by $(v_t f)(x)= e^{it}f(x)$ for  $f\in\hg$.  We use
Proposition \ref{p:4.10}, with 
$a=m_\infty$ and $b=\omega$. Conditions \eqref{e:4.19} and
\eqref{e:4.21} are trivially satisfied. Condition \eqref{e:4.22}
follows from Lemma \ref{l:6.2}. Therefore, $H_\lambda\in
\Cc^1\big(M_\infty;\Dc\big( |H_\lambda|^{1/2}\big),  \Dc \big( |H_\lambda|
^{1/2}\big)^*\big)$ and thus  \cite{ABG}[Lemma 7.5.3] gives $H_\lambda\in
\Cc^1(M_\infty)$.  Therefore, Proposition \ref{p:ntau} gives that
$\Dc(H_\lambda)\cap \Dc(M_\infty) = \Dc(N)\cap\un_\Kr\!\otimes
\Dc\big(\dG(\omega)\big)=\Dc(B_\infty)$. This is an obvious
core for $M_\infty$.  

Now, Lemma \ref{l:avoid} implies point ii) and also gives 
the statements on $\Dr=\Dr_\infty$ in iii). By Proposition
\ref{p:ntau} and {\bf (I1a)}, we have that $S_\infty$ is $H_0$-form
bounded. Then, the norm $\|\cdot \|_\Gr$, given by \eqref{e:G}, is
equivalent  to $\sqrt{\langle \cdot   , (M_\infty+H_0 +1)\,
  \cdot\rangle }$ on $\Dr$. Since $\Dr$ is a form core for $B_\infty$,
we infer $\Gr=\Gr_\infty$. \qed 

From now on, we drop the subscripts for $\Dr$ and $\Gr$. We clarify
the $\Cc^1$ property. The hypothesis {\bf (M2)} is checked in Theorem
\ref{t:ms}.  

\begin{lemma}\label{l:6.6.0}
Assume 	{\bf (I0)} and {\bf (I1a)}. Then, 

{\rm i)} $\{W_{\infty,t}\}_{t\geq 0}$ $b$-stabilizes $\Gr$ and $\Gr^*$. 

{\rm ii)} $H_\lambda\in\Cc^1( A_\infty; \Gr, \Gr^*)$ and
$[H_\lambda, i A]= H'_\lambda$ on $\Dr$. 
\\
Therefore, hypotheses {\bf (M3)} and  {\bf (M4)} are fulfilled. 
\end{lemma}
\proof We apply Proposition \ref{p:4.10}. As in the proof of Proposition
\ref{p:ess}, we work in polar coordinate through the isomorphism
\eqref{e:T}. In this representation, the operator $b$ acts by $\tilde
b= (r^2+1)^{1/2}\otimes \un$ in $\tilde \hg$. Using \eqref{e:wtilde}
and \eqref{e:wadj}, we obtain  
\begin{eqnarray*}
\tilde w_{\infty,t}^* \tilde b \tilde w_{\infty,t}=
b\big(\phi_{\infty, t}(\cdot)\big) \mbox{ and } \tilde w_{\infty,t} \tilde b
\tilde w_{\infty,t}^*= \un_{\R^+}\big(\phi_{\infty, -t}(\cdot)\big)
b\big(\phi_{\infty, -t}(\cdot)\big).
\end{eqnarray*} 
Therein, the flow $\phi_{\infty,t}$ was extended in $\R$. We have, 
\begin{eqnarray}\label{e:6.6}
\hspace{1cm}|b\big(\phi_{\infty, t}(r)\big)- b(r)|\leq \|\nabla b\|_\infty
|\phi_{\infty, t}(r)-r|\leq \|\nabla b\|_\infty |t|, \mbox{ for }
0\leq |t|\leq 1.  
\end{eqnarray}  
We infer 
$1\leq b\big(\phi_{\infty, t}(r)\big) \leq \|\nabla b\|_\infty \big(1+|t|\big)
b(r)$, for  $0\leq |t|\leq 1$.  
Hence, the condition \eqref{e:4.19} is satisfied. The $C_0$-semigroup
$\{W_{\infty,t}\}_{t\in\R^+}$ and $\{W_{\infty, t}^*\}_{t\in\R^+}$
$b$-stabilizes $\Gr$.  
  
We prove the second point with the help of Proposition \ref{p:4.10}
ii) and iii). First, $\omega\leq b$. Now, $\omega w_{\infty,
  t}-w_{\infty, t}\omega= \big(\omega -
\omega\big(\phi_{\infty,-t}(\cdot)\big)\big)w_t$. By \eqref{e:6.6}, we
obtain that $|\phi_{\infty, t}(r)-r|\leq C |t| b(r)$ and hence 
$|\omega -
\omega\big(\phi_{\infty,-t}(\cdot)\big)|\leq C|t|b$, for all
$t\in[0,1]$. Since $\{\omega_t\}_{t\in \R^+}$ $b$-stabilizes
$\Dc(b^{1/2})$, we get \eqref{e:4.21}. Now by Lemma \ref{l:6.2}, we
check \eqref{e:4.22}. We obtain $H_\lambda\in \Cc^1(A_\infty; \Gr,
\Gr^*)$.  \qed

\subsection{Estimation on the first perturbed commutator}\label{s:1comm'}
We now add the finite rank perturbation $B_\varepsilon$ to the conjugate
operator. We consider the conjugate operator $\hat A_n$, given by
\eqref{e:Ahat}. We denote with a hat the perturbed operators. Set 
\begin{eqnarray}\label{e:M&R'}
 \hat S_n:= S_n+ [H_\lambda, i \lambda \theta B_\varepsilon], \quad
 \mbox{ for all } n\in \N^*\cup\{\infty\}. 
\end{eqnarray} 
Note that the operator $M_n$, given in \eqref{e:M&R}, is unaffected by
$B_\varepsilon$.     

Although $B_\varepsilon$ is a finite rank perturbation, one needs to
be careful, especially in the $2$-commutators properties. We give the
key-lemma which allows us to transfer safely properties  of $A_n$ to
$\hat  A_n$. We point out that Lemma \ref{l:commu} shows that
$[H_\lambda, B_\varepsilon]$ is also a finite rank operator in
$\Hr$. Recall that $\Gr=\Gr_\infty$ is given in \eqref{e:B}. 
\begin{lemma}\label{l:finite}
Assume 	{\bf (I0)}. We have:

{\rm i)}  $B_\varepsilon$ is a finite  rank self-adjoint     operator. 

{\rm ii)} $B_\varepsilon\in \Bc(\Gr)$. 

{\rm iii)} Assume also {\bf (I1a)}, then $B_\varepsilon$ is belonging to $\Cc^1(A_\infty; \Gr, \Gr)$.
\end{lemma} 

\proof Since $P$ is of finite rank and $B_\varepsilon$ is symmetric,
we need only to show that $B_\varepsilon$ is bounded. 
{\bf (I0)} gives that
$P\phi(\alpha)\overline P= P\alpha\overline P$ belongs to $\Bc(\Hr,
\Kr\!\otimes \Dc(\omega^{-1/2}))$. Now, recall that $\varepsilon
R_\varepsilon^2 = \im (H_0-k+ i \varepsilon )^{-1}$ and that $1\otimes
\omega^{1/2} (H_0-k \pm i \varepsilon )^{-1}$ is bounded by
functional calculus in $\Kr\!\otimes \hg$. This concludes i). 

For point ii), note that $B_\varepsilon\in \Bc(\Gr)$ is equivalent to
$B_\varepsilon \in \Bc\big(\Dc(|H_0|^{1/2})\big)$, since
$P\alpha\overline P$ is with image in the $1$-particle space. 
Hence, the assertion follows by noticing that 
$1\otimes \omega^{1/2}(1+\omega)^{1/2}(H_0-k\pm i \varepsilon)^{-1}$
is bounded in $\Kr\!\otimes \hg$. 

As in ii), it is enough to show that $T:=P \phi(\alpha)\overline
P(H_0 -z )^{-1}$ and  its adjoint are in $\Cc^1\big(A_\infty;
\Dc(|H_0|^{1/2}), \Dc(|H_0|^{1/2})\big)$, where $z\in\C\setminus \R$.   
We treat $T$. Note that $H_0|_{\Kr\!\otimes \hg}\in \Cc^1(A_\infty)$. 
Using \eqref{e:T}, we have:
\begin{align*}
[T, iA_\infty]= P (1\otimes \partial_r) \alpha \overline P(H_0 -z
)^{-1}-  P \alpha \overline P(H_0 -z )^{-2}.
\end{align*} 
Like in ii), the second term is easily bounded in
$\Dc(|H_0|^{1/2})$. The boundedness of the first one is ensured
by the second part  of {\bf (I1a)}. \qed

As an immediate corollary, we infer from Lemma \ref{l:6.4} the following. 

\begin{lemma}\label{l:6.4'}
Assume {\bf (I0)} and {\bf (I1a)}. Then:

{\rm i)} $H_\lambda\in\Cc^1(M_\infty)$, $\Dc(H_\lambda)\cap
\Dc(M_\infty)$ is a core for $M_\infty$, $\hat
S_\infty$ is  symmetric and lies in $\Bc(\Dc(H_0), \Hr)$. 

{\rm ii)} Let $\hat H'_\lambda$ be the closure of 	$M_\infty+\hat
S_\infty$ defined on   	$\Dc(H_\lambda)\cap
\Dc(M_\infty)$. Therefore, $H_\lambda$ and 
$\hat H'_\lambda$ satisfy {\bf (M1)}. 

{\rm iii)} $\Dr=\Dc(\hat H'_\lambda)\cap \Dc(H_\lambda)
=\Dc(M_\infty)\cap \Dc(H_\lambda)$ and $\Gr$ is the same as in \eqref{e:G}. 
\end{lemma}

We now strengthen Lemma \ref{l:6.6.0} and check {\bf (M3)} and  {\bf
  (M4)}. The hypothesis {\bf (M2)} is  checked in Theorem \ref{t:msFGR}. 

\begin{lemma}\label{l:6.6}
Assume 	{\bf (I0)} and {\bf (I1a)}. Then,

{\rm i)} $\{\widehat W_{\infty,t}\}_{t\geq 0}$ $b$-stabilizes     $\Gr$ and $\Gr^*$. 

{\rm ii)} $\Cc^1(A_\infty, \Gr, \Gr^*)= \Cc^1(\hat
A_\infty, \Gr, \Gr^*)$. 

{\rm iii)} $H_\lambda\in\Cc^1(\hat A_\infty; \Gr, \Gr^*)$ and
 	$[H_\lambda, i \hat A]=  \hat H'_\lambda$ on $\Dr$.
\\
Therefore, hypotheses {\bf (M3)} and  {\bf (M4)} are fulfilled. 
\end{lemma}
\proof We consider $\{\widehat W_{\infty,t}\}_{t\in\R}$. The argument is
the same for the adjoint.   Let $A_\infty'$ be the generator of
$\{W_{\infty,t}\}_{t\in\R}$ in $\Gr$. As in
\eqref{e:Ahat}, set $\widehat{A_\infty}':= A_\infty'+ \lambda\theta
B_\varepsilon$. Thanks to Proposition \ref{p:kato}, since
$B_\varepsilon\in \Bc(\Gr)$, $\widehat{A_\infty}'$ is the
generator of a $C_0$-semigroup in $\Gr$. We name it
$\{\widehat W_{\infty,t}'\}_{t\in\R}$. By duality and interpolation,
it extends to a $C_0$-semigroup in $\Hr$. Comparing the generators, we
obtain that $\{\widehat W_{\infty,t}'\}_{t\in\R}$ is really the
restriction of $\{\widehat W_{\infty,t}\}_{t\in\R}$ and it gives point
i). By Lemma \ref{l:6.6.0}, it is enough to show ii) to get iii). 
Proposition \ref{p:2comm} and the boundedness of 
$B_\varepsilon$ in $\Gr$ and $\Gr^*$ give the former.\qed

\subsection{The Virial theorem}\label{s:virial}
In order to obtain a Virial theorem, we proceed like in \cite{GGM2,
Ski} by approximating the conjugate operator. Indeed, since $\hat
H'_\lambda$ is not $H_\lambda$-bounded, one can not apply \emph{a
priori} $H'_\lambda$ to an eigenfunction of $H_\lambda$ even in the
form sense. In this section, we use the hypotheses  {\bf (I0)} and
{\bf (I1)}. Here, {\bf   (I1)} means {\bf (I1a)} and {\bf (I1b)}. 
In a zero temperature setting, this method is less demanding in
hypotheses than the one used in \cite{Merkli}, see for instance
\cite{Merkli}[Proposition 6.1]. Note that we do not deal with the
positive temperature Hamiltonians treated therein.

\begin{lemma}\label{l:unif}
Assume {\bf (I0)} and {\bf (I1)}. Then $\phi(ia_n \alpha)$ tends to
$\phi(ia_\infty \alpha)$, as quadratic forms on
$\Dc(|H_\lambda|^{1/2})$, as $n$ goes to infinity. 
\end{lemma}
\proof Thanks to Proposition \ref{p:ntau}, it is enough to show that
$\|(a_n-a_\infty)\alpha\|_{\Bc(\Kr, \Kr\!\otimes \hg)}$ and that
$\|\omega^{-1/2}(a_n-a_\infty)\alpha\|_{\Bc(\Kr, \Kr\!\otimes 
\hg)}$ tend to $0$ as $n$ goes to infinity. 

We start with the first point. Like in the proof of Proposition
\ref{p:ess}, we work in polar coordinates. We focus on the expression
of $\tilde a_n$ obtained in \eqref{e:atilde}.  We have $\alpha \in \Bc(\Kr,
\Kr\!\otimes \dot\Hr^1\otimes L^2)$. Moreover, since $m_n(r)\leq
m_\infty(r)=1$ and $m_n$ converges simply to $1$ almost everywhere, by
the Lebesgue dominated convergence theorem,  we obtain $\|(m_n -
m_\infty) \partial \alpha\|_{\Bc(\Kr, \Kr\!\otimes \hg)}$ tends to
$0$. We treat the term in $\big(m_n'(r)-m_\infty'(r)\big) \alpha =
m_n'(r)\alpha = m_n'(r)r^{a} r^{-a}\alpha$. As $a>1/2$, 
dominated convergence proves it tends to $0$ in $\Bc(\Kr, \Kr\!\otimes
\hg)$. The proof of the second point is the same but use the fact that
$r^{-a}\alpha \in \Bc(\Kr, \Kr\!\otimes \hg)$ for the term in
$m_n'$, for some $a>1$. \qed 

We point out that if one knows that $\omega^{-1}\alpha\in\Bc(\Kr, \Kr\!\otimes
\Cc_0(\R^+)\otimes L^2(S^{d-1}))$, one may relax {\bf (I1b)} and take
$a=1$. Here $\Cc_0(\R^+)$ denotes the continuous functions vanishing
in $0$ and in $+\infty$.    

\begin{lemma}\label{l:6.3.0}
Assume $n$ finite, {\bf (I0)} and {\bf (I1a)}. Then,
$\{\widehat W_{n,t}\}_{t\in\R}$ $b$-stabilizes the form domain
of $H_\lambda$. 
\end{lemma}
\proof First we apply Proposition \ref{p:4.10} i) with $v_t=w_{n,t}$ and $b=w$. 
As we have a $C_0$-group, by taking $t$ negative 
we obtain the result for the adjoint. As in the proof of Proposition
\ref{p:ess}, we denote by $\phi_{n,t}:\R^d\rightarrow \R^d$ the flow
generated by the smooth vector field $\overrightarrow{s_n}$.  Since
$m_n(0)=0$, we have 
\begin{align}
\nonumber
|\phi_{n,t}(k)-k|&= |\phi_{n,t}(k)-\phi_{n,0}(k)|\leq \int_0^{|t|}
\left|m_n(\phi_{n,s}(k))- m_n(0)\right|\, ds 
\\ 
\label{e:pourGron}
&\leq \|\nabla m_n\|_{\infty} \int_0^{|t|} |\phi_{n,s}(k)| \, ds,
\mbox{ for all } t\in \R. 
\end{align}
By the Gronwall lemma, we infer there is $C$ such that
$|\phi_{n,t}(k)|\leq C|k|$, for all $t\in[1,1]$. Plugging back into
\eqref{e:pourGron}, we obtain $|\phi_{n,t}(k) - k|\leq C|tk|$, for all
$t\in[1,1]$. Now using \eqref{e:wtilde} and \eqref{e:wadj}, we infer
$e^{-ita_n} w e^{ita_n}= w\big(\phi_{n,t}(\cdot) \big)$. Since $m_n$
is globally Lipschitz, there is $C'$ such that  
\begin{eqnarray}\label{e:Gronw}
|w(\phi_{n,t}(k))- w(k)|\leq C'|t|w(k), \mbox{ for all } t\in[1,1].
\end{eqnarray} 
Hence, we satisfy the hypothesis \eqref{e:4.19} and
$\Dc(|H_\lambda|^{1/2})$ is $b$-stable under $\{W_{n,t}\}_{t\in\R}$. 

We now take care about $\{\widehat W_{n,t}\}_{t\in\R}$. Let $A_n'$ be the
generator of $\{W_{n,t}\}_{t\in\R}$ in  $\Dc(|H_\lambda|^{1/2})$. As
in \eqref{e:Ahat}, set $\widehat{A_n}':= A_n'+ \lambda\theta
B_\varepsilon$. By Lemma \ref{l:finite} ii) and the fact
that $B_\varepsilon $ is with values in the $0$ and $1$ particles
space, we get $B_\varepsilon$  bounded in $\Dc(|H_\lambda|^{1/2})$. 
Thanks to Proposition \ref{p:kato},  $\widehat{A_n}'$ is the generator
of a $C_0$-group in $\Dc(|H_\lambda|^{1/2})$. We name it $\{\widehat
W_{n,t}'\}_{t\in\R}$. By duality and interpolation, it extends to a
$C_0$-group in $\Hr$. Comparing the generators, we obtain that
$\{\widehat W_{n,t}'\}_{t\in\R}$ is really the restriction of
$\{\widehat W_{n,t}\}_{t\in\R}$ and this gives the result.\qed 

\begin{lemma}\label{l:6.3.1}
Assume $n$ finite, {\bf (I0)} and {\bf (I1a)}. Then
$H_\lambda\in\Cc^1(\hat A_n)$. Moreover:  
\begin{eqnarray}\label{e:commun}
\big[H_\lambda, i\hat A_n\big]= M_n+\hat S_n,
\end{eqnarray}
holds true in the sense of forms on $\Dc(|H_\lambda|^{1/2})$.
\end{lemma}
\proof Using again \eqref{e:Gronw}, we check \eqref{e:4.21}. We get
$[H_0, iA_n]^\circ=  \un_\Kr\!\otimes \dG([\omega, ia_n]^\circ)$ in the
sense of form on $\Dc(|H_\lambda|^{1/2})$. By computing $[\omega,
  ia_n]^\circ$ on the core $\Cc^\infty_c(\R^d\setminus\{0\})$, we
obtain $[\omega, ia_n]^\circ= m_n$. Now, by Lemma \ref{l:6.2}, we can
use Proposition \ref{p:4.10} iii) and deduce  $\big[H_\lambda, i
  A_n\big]= M_n+ S_n$ in the sense of forms on
$\Dc(|H_\lambda|^{1/2})$. Finally, by Lemma \ref{l:commu}, $[H_\lambda,
B_\varepsilon]$ is of finite rank, we also obtain \eqref{e:commun} on
the same domain.  

Now, $H_\lambda\in\Cc^1\big(\hat A_n;
\Dc\big(|H_\lambda|^{1/2}\big), \Dc\big( |H_\lambda|^{1/2}
\big)^*\big)$ by Lemma \ref{l:6.3.0} and Proposition \ref{p:2comm}. 
We apply \cite{GGM2}[Lemma 6.3] to get
$H_\lambda\in\Cc^1(\hat A_n)$. \qed  

Therefore, the Virial theorem holds true when $\hat A_n$ is the
conjugate operator and when $n$ is finite. However, there is no Mourre
estimate for $\hat A_n$ but only one for $\hat A_\infty$. To 
overcome this problem, we take advantage of the monotone convergence
of $[H_0,i A_n]$ to $[H_0,i A_\infty]$ and of the uniformity given in Lemma
\ref{l:unif} to prove:  
\begin{proposition}[Virial theorem]\label{p:6.5}
Assume 	{\bf (I0)} and {\bf (I1)}. Let $u$ be an
eigenfunction of $H_\lambda$ then $u\in \Dc( N^{1/2})$ and $\langle u,
(M_\infty+ \hat S_\infty) u\rangle=0$, as a quadratic form on
$\Dc(N^{1/2})\cap\Dc(H_\lambda)$.  
\end{proposition}
\proof 
First, $M_n$ is a bounded form for $H_\lambda$. Note that
$0\leq m_n\leq m$ implies $0\leq \dG(m_n)\leq  \dG(m)$ for all
$n$. Now, since $m_n$ is increasing and converges to $m$ as $n$ goes
to infinity, monotone convergence gives 
\begin{eqnarray*}
0\leq \langle g, M_n g  \rangle\leq  
\langle g, M_\infty g  \rangle 
\mbox{ and }
\langle g, M_n g  \rangle \build{\longrightarrow}_{n\rightarrow \infty}^{}
\langle g, M_\infty g  \rangle,
\end{eqnarray*} 
for all $g\in \Dc(M_\infty)\cap \Dc(H_\lambda)$. Using
some Cauchy sequences, this holds true also in the sense of forms for $g\in
\Dc(M_\infty^{1/2})\cap \Dc(H_\lambda)$. By authorizing the value
$+\infty$ on the two r.h.s.\  when $g\notin\Dc(M_\infty^{1/2})$, one allows 
$g\in\Dc(H_\lambda)$. On the other hand, Lemma
\ref{l:6.2} gives that  $\hat S_n$ tends to $\hat S_\infty$  
as a quadratic form on $\Dc(H)$.  

Let $\dot H$ be the closure of quadratic form $\langle u, \hat
H'_\lambda u \rangle $ defined on $\Dc(M_\infty)\cap \Dc(H)$. It is
given by the quadratic form  $\langle u, (M_\infty + \hat S_\infty)
u\rangle$ defined on $\Dc(M^{1/2}_\infty)\cap \Dc(H)$. Take now an
eigenfunction $u$ of $H_\lambda$. By Lemma \ref{l:6.3.1} and the
Virial theorem, see \cite{ABG}[Proposition 7.2.10], we get $\langle u,
(M_n+\hat S_n) u \rangle=0$. By letting $n$ go to infinity and
noticing that $\Dc(M_\infty^{1/2})=\Kr\!\otimes \Dc(N^{1/2})$, we get
the result. \qed  
\subsection{Estimation on the second commutator}\label{s:2comm}
In this section, we discuss the second commutator hypothesis 
{\bf (I2)} so as to obtain a limiting absorption principle through the
Theorem \ref{t:LAP}. 
We stress we forgo the hypothesis {\bf (I1b)} in this
section. We start with the important remark.

\begin{lemma}\label{l:changeregu}
We have $\Cc^2(A_\infty, \Gr, \Gr^*)= \Cc^2(\hat
A_\infty, \Gr, \Gr^*)$. 
\end{lemma} 
\proof It is enough to
show one inclusion. Using Proposition \ref{p:2comm} and the invariance
of $\Gr$ and $\Gr^*$ given in Lemmata \ref{l:6.6.0} and \ref{l:6.6},
one may work directly with $A_\infty$ and $\hat A_\infty$.  Let $H\in
\Bc(\Gr, \Gr^*)$ be in $\Cc^2(A_\infty, \Gr, \Gr^*)$. One justifies
the next expansion, by working in the form sense on
$\Dc\big((A_\infty^*)^2|_{\Gr}\big)\times \Dc\big((A_\infty)^2|_{\Gr}\big)$. This is
legal by using Lemma \ref{l:finite}. We have: 
\begin{align*}
[[H, \hat A_\infty], \hat A_\infty]=&
[[H, A_\infty], A_\infty] +  [[H, A_\infty], \lambda\theta
  B_\varepsilon] \\
& [[H, \lambda \theta   B_\varepsilon], A_\infty] + [[H, \lambda
    \theta   B_\varepsilon] , \lambda     \theta   B_\varepsilon].
\end{align*} 
The first term is in $\Bc(\Gr, \Gr^*)$ by hypothesis. 
For the second one, note that $[H, A_\infty]\in\Bc(\Gr, \Gr^*)$ since
$H$ is $\Cc^1(A_\infty, \Gr, \Gr^*)$. 
For the third one, we expand the
commutator inside, use again that $H\in\Cc^1(A_\infty, \Gr, \Gr^*)$
and finish with Lemma \ref{l:finite} iii).
For the last one, one expands it
and use Lemma \ref{l:finite} ii).  \qed

We start by discussing the $\Cc^2$ theory used in \cite{GGM2, Ski}
and check the point {\bf   (M5')}. Through the isomorphism given by
\eqref{e:T}, we suppose the stronger
\begin{itemize}
\item[{\bf (I2')}] $\alpha\in \Bc\big(\Kr, \Kr\!\otimes
  \dot\Hr^{2}(\R^+)\otimes L^2(S^{d-1})\big)$.
\end{itemize}
This hypothesis is stronger than $\alpha\in \Bc\big(\Kr, \Kr\!\otimes
\dot\Hr^{s}(\R^+)\otimes L^2(S^{d-1})\big)$ for $s>1$, the one used
in \cite{DJ}[Theorem 6.3]. 

\begin{lemma}\label{l:6.7}
Assume {\bf (I0)}, {\bf (I1a)} and {\bf (I2')}. Then
$H_\lambda\in\Cc^2(\hat A_\infty, \Gr, \Gr^*)$ and 
\begin{eqnarray*}
[\hat H'_\lambda, i \hat A_\infty]=  \lambda \phi\big(a^2_\infty \alpha\big)
+\lambda\theta [[H_\lambda, B_\varepsilon ], iA] + \lambda^2\theta^2
[\hat H_\lambda', B_\varepsilon ]. 
\end{eqnarray*}
Therefore, the hypothesis {\bf (M5')} is fulfilled.
\end{lemma}
\proof We use  Proposition \ref{p:4.10} ii) and iii)
for the operator $H:=N - \lambda \phi(i a_\infty
\alpha)$. Point ii) is trivially satisfied. 
The hypothesis {\bf (I2)} and Proposition \ref{p:ntau} give
\eqref{e:4.22}. We obtain $H\in \Cc^1(A_\infty; \Gr, \Gr^*)$. \qed

We now work with the hypothesis {\bf (I2)} which is weaker than the
one used in \cite{DJ}. Thanks to Lemma \ref{l:changeregu}, we have
\begin{align*}
\Cc^{1,1}(A_\infty, \Gr, \Gr^*) &:= \big(\Cc^{2}(A_\infty, \Gr, \Gr^*),  \Bc(\Gr,
\Gr^*)\big)_{1/2, 1}
\\
&= \big(\Cc^{2}(\hat A_\infty, \Gr, \Gr^*),  \Bc(\Gr,
\Gr^*)\big)_{1/2, 1}=: \Cc^{1,1}(\hat A_\infty, \Gr, \Gr^*).
\end{align*} 
We refer to \cite{ABG, Tri} for real interpolation. 
We obtain:

\begin{lemma}\label{l:6.7'}
Assume {\bf (I0)}, {\bf (I1a)} and {\bf (I2)}. Then
$H_\lambda\in\Cc^{1,1}(\hat A_\infty, \Gr, \Gr^*)$ and 
the hypothesis {\bf (M5)} is fulfilled.
\end{lemma}
\proof By Lemma \ref{l:6.7}, we have $H_0\in \Cc^2(\hat A_\infty, \Gr,
\Gr^*)$. It is enough to show that $\phi(\alpha)\in\Cc^{1,1}(A_\infty, \Gr,
\Hr)$. By \cite{DJ}[Lemma 2.7], we have $W_{\infty,  t} \phi(\alpha)=
\phi(w_{\infty, t}\alpha) W_{\infty, t}$ for $t\geq 
0$. By Proposition \ref{p:ntau} and  $b\geq 1$ and since $\{W_{\infty,
t}\}$ $b$-preserves $\Gr$, we get
\begin{align*}
\int_0^1\big\|[W_{\infty,  t}[W_{\infty,  t}, \phi(\alpha)]]\big\|_{\Bc(\Gr,
\Hr)}\, \frac{dt}{t^2}& \leq 
\int_0^1\big\|\phi([w_{\infty, t}[w_{\infty,  t}, \alpha]])
W_{\infty, 2t}\big\|_{\Bc(\Gr, \Hr)}\, \frac{dt}{t^2} 
\\
&
\leq C \int_0^1\big\|[w_{\infty, t}[w_{\infty,  t}, \alpha]]
\big\|_{\Bc(\Kr, \Kr\!\otimes \hg)}\, \frac{dt}{t^2}. 
\end{align*} 
The latter is finite if and only if 
$\alpha$ belongs to $\big(\Bc\big(\Kr, \Dc(a_\infty^2)\big), \Bc(\Kr, 
\Kr\!\otimes \hg)\big)_{1/2,   1}$.  On the other hand, using the
isomorphism \eqref{e:T} and Proposition \ref{p:ess}, this space is the 
same as $\big( \Bc\big(\Kr, \Kr\!\otimes \dot\Hr^2(\R^+)\otimes
L^2(S^{d-1})\big), \Bc(\Kr, \Kr\!\otimes \tilde\hg)\big)_{1/2,
  1}$. Finally, using \cite{Tri}[Section 2.10.4], this is equivalent
to the fact that $\alpha$ satisfies {\bf (I2)}. \qed 

\section{A Mourre estimate far from the thresholds}\label{s:far}
\setcounter{equation}{0}
\subsection{The result} The aim in this part is to show a Mourre
estimate far from thresholds for small coupling constants. This is a
well-known result, see \cite{BFS,   DJ} for instance.  For the sake of
completeness, we give a proof of the estimate. Doing
so, we point out, in Remark \ref{r:motiv}, where the lack of positivity
occurs above the thresholds. We use the approach based on the
theory described in Appendix \ref{s:ms}. To obtain
information just above the thresholds and without supposing the Fermi
golden rule, one should add a compact term in \eqref{e:ms}, see
\cite{GGM, Ski}.

\begin{theorem}\label{t:ms}
Let $\Ic_0$ be a compact interval containing no element of $\sigma(K)$. 
Suppose also that {\bf(I0)} and {\bf(I1a)} are satisfied. Then, for
all open interval $\Ic\subset \Ic_0$:

{\rm i)} There are $M_\infty\geq 1$ and $S_\infty$ a
$|H_\lambda|^{1/2}$-bounded operator such that $[H_\lambda, i
  A_\infty]= M_\infty+S_\infty$ holds in the sense of forms on
$\Dc(N^{1/2})$.   

{\rm ii)} The conditions {\bf(M1)}--{\bf(M4)} are satisfied. 

{\rm iii)} There is $\lambda_0>0$ such that the
  following extended Mourre estimate  
\begin{eqnarray}\label{e:ms}
M_\infty + S_\infty \geq a(\lambda) E_\Ic(H_\lambda) - b(\lambda)
E_{\Ic^c}(H_\lambda)\langle H_\lambda \rangle.  
\end{eqnarray} 
holds true in the sense of forms on $\Dc(N^{1/2})$,  for 
all $|\lambda| \leq \lambda_0$. Here, $a(\lambda)$ is positive and can
be written as $\big(1+O(\lambda)\big)$. Besides, $b(\lambda)$ is also
positive. 
  
{\rm iv)} If {\bf (I1b)} holds true, then $H_\lambda$ has no
  eigenvalue in $\Ic$, for all $|\lambda| \leq \lambda_0$. 
  
{\rm v)} If {\bf (I2)} holds true (and not necessarily  {\bf (I1b)}),
then $H_\lambda$ has no  eigenvalue in the interior of $\Ic$, for all
$|\lambda| \leq \lambda_0$. Moreover, one obtains the estimations of
the resolvent given in Theorem \ref{t:intro1}.  
\end{theorem} 

\proof By Lemma \ref{l:6.4}, we have the first point. The point ii) is
shown in Section \ref{s:1comm}.  The point iii) follows from
Proposition \ref{p:ms1}. Indeed, since $S_\infty$ is form bounded with
respect to $H_\lambda$, we have that for all $\eta>0$
\begin{align}\nonumber
 E_\Ic(H_\lambda) S_\infty E_{\Ic^c}(H_\lambda) +E_{\Ic^c}(H_\lambda)
 S_\infty E_{\Ic}(H_\lambda)  \geq &
\\\label{e:eta}
&\hspace{-5cm} -\eta E_\Ic(H_\lambda) S_\infty \langle H_\lambda \rangle^{-1} S_\infty
 E_\Ic(H_\lambda) - \eta^{-1} E_{\Ic^c}(H_\lambda) \langle H_\lambda \rangle.
\end{align} 
The point iv) follows from the Virial Theorem, Proposition
\ref{p:6.5}. Finally, Theorem \ref{t:LAP} gives point v),
the space $\Gr$ appearing therein is identified in
Lemma \ref{l:6.4}. \qed

\subsection{The inequality}
Here we establish the extended Mourre estimate away from the
threshold. We use only {\bf (I0)} and  {\bf (I1a)} and do not assume
any Fermi golden rule assumption.  

\begin{proposition}\label{p:ms1}
Let $\Ic_0$ be a compact interval such that $\sigma(K)\cap
\Ic_0=\emptyset$. Let $\Ic$ be an open interval included in $\Ic_0$. Let
$M_\infty:=N+ \un \otimes P_\Omega \geq 1$ and let
$S_\infty:=-\un\otimes P_\Omega- \lambda\phi(ia_\infty \alpha)$.  For
$\lambda$ small enough, we get   
\begin{eqnarray}\label{e:ms1}
M_\infty+E_\Ic(H_\lambda) S_\infty E_\Ic(H_\lambda) \geq
\big(1+O(\lambda)\big)E_\Ic(H_\lambda), 
\end{eqnarray}
holds true in the sense of forms on
$\Dc(N^{1/2})$.
\end{proposition} 
\proof  Let $\Jc$ be a compact set containing $\Ic$ 
and contained in the interior of $\Ic_0$. Note that \eqref{e:H0p} gives
$E_\Jc(H_0) \un_\Kr\! \otimes P_\Omega = 0$. By Proposition
\ref{p:ntau}, we derive: 
\begin{eqnarray}\label{e:H0p'}
E_\Jc(H_0) S_\infty E_\Jc(H_0) = \lambda  E_\Jc(H_0) \phi(ia_\infty \alpha)
E_\Jc(H_0) = O(\lambda)E_\Jc(H_0). 
\end{eqnarray}  
As $M_\infty\geq 1$, it remains to prove that $E_\Ic(H_\lambda)
S_\infty E_\Ic(H_\lambda)=O(\lambda)E_\Ic(H_\lambda)$. 
We insert $E_\Jc(H_0)+E_{\Jc^c}(H_0)$ on the right and on the left of $S_\infty$.
By \eqref{e:H0p'}, all the four terms are actually
$O(\lambda)E_\Ic(H_\lambda)$. Indeed, Proposition \ref{p:ntau} gives
for instance that
\begin{align*}
E_\Ic(H_\lambda)E_\Jc^c(H_0)  S_\infty E_\Jc(H_0)E_\Ic(H_\lambda) =
O(\lambda) E_\Ic(H_\lambda). 
\end{align*} 
For the right hand side, take $h\in\Cc^\infty_c(\Jc)$ so that $h|_{\Ic}=1$.
We have 
\begin{align*}
E_\Ic(H_\lambda)E_{\Jc^c}(H_0)= E_\Ic(H_\lambda)\big(h(H_\lambda)-h(H_0)\big)
E_{\Jc^c}(H_0)=O(\lambda),
\end{align*} 
by Lemma \ref{l:HS}. \qed

\begin{rem}\label{r:motiv}
This proof would not work over one of thresholds $\{k_i\}_{i=0,\ldots,
  n}$. Here, we use  in a drastic way that $E_\Jc(H_0) \un \otimes
P_\Omega=0$. However, when  $\sigma(K)\cap \Ic= \{k_i \}$, this
expression is never $0$ and is of norm $1$. A brutal estimation would
give  
\begin{eqnarray}
M+E_\Ic(H_\lambda) S_\infty E_\Ic(H_\lambda) \geq O(\lambda)E_\Ic(H_\lambda).
\end{eqnarray}
We have no control on the sign. This is no surprise as we
know that one may uncouple the two parts of the  system and an
eigenvalue can remain, see Section \ref{s:thre}. To control the sign,
one  needs to gain some positivity just above $P_{k_i}\otimes
P_\Omega$. This would be the r\^ole of the Fermi golden rule and of
the operator $B_\varepsilon$.  
\end{rem}

Here we have used the elementary:

\begin{lemma}\label{l:HS}
Let $h\in\Cc^\infty_c(\R)$ and $s\leq 1/2$. Let $V$ be symmetric
operator being $H_0$-form bounded operator, with constant lower than
$1$. Then, there is $C$ such that   
\begin{eqnarray*}
\big\|\langle H_0\rangle^s\big(h(H_0) -
h(H_0+\lambda V)\big)\big\|\leq C |\lambda|.   
\end{eqnarray*} 
\end{lemma} 

\section{A Mourre estimate at the thresholds}\label{s:thre}
\setcounter{equation}{0}

In this section we would like to study the absence of eigenvalue above
one of the thresholds.  From a physical point of view, as soon
as the interaction is on, one expects the embedded eigenvalues to
disappear into the complex plane and to turn into resonances. This is
however not mathematically true as one may uncouple the Bosonic Field
and the atom. Take for instance $\omega$ bounded,  $\alpha\in\Bc(\Kr,
\Kr\!\otimes \hg)$, given  by $\alpha(x):= 1\otimes b$, for all $x\in
\Kr$ and where $\omega b\in \hg$. After a dressing transformation, see
for instance \cite{D}[Theorem 3.5], the operator $H_\lambda$ is
unitarly equivalent to the free operator $K\otimes \un_{\Gamma(\hg)}+
\un_\Kr\!\otimes \dG(\tilde \omega_\lambda)$, for some $\tilde
\omega_\lambda \in \Bc(\hg)$. Therefore, $H_\lambda$ has the same
eigenvalues as $H_0$ for all $\lambda$. Note that this is no
restriction to suppose that $\omega$ is bounded thanks to the
exponential law, see for instance \cite{BSZ}[Section 3.2]. We couple
the two systems through a Fermi golden rule assumption.    

\subsection{The Fermi golden rule hypothesis}\label{s:FGR}
We choose one eigenvalue $k_{i_0}$ of $H_{\rm{el}}$ for $i_0\pg
0$. Let $P:=P_{k_{i_0}}\otimes P_{\Omega}$ and let $\overline{P}:=1-P$. Note
that $P$ is of finite rank. We give an implicit hypothesis
on $\alpha$ and explain how to check it in Appendix
\ref{s:level}.  \begin{definition} We say that the \emph{Fermi golden
    rule} holds true at energy $k$ for a couple $(H_0, \alpha)$ if
  there exist positive $\varepsilon_0$, $c_1$ and $c_2$ such that  
\begin{eqnarray}\label{e:FGR0}
c_1 P\geq P\phi(\alpha)\overline{P}\,\im(H_0-k +i\varepsilon)^{-1}
\overline{P}\phi(\alpha) P\geq c_2 P, 
\end{eqnarray} 
holds true in the sense of forms, for all $\varepsilon_0 >\varepsilon>0$.
\end{definition}
Due to the Fock space structure, one may omit $\overline P$ in
\eqref{e:FGR0} but we keep it to emphasize the link between 
hypotheses of this type in other fields (like for Schr\"odinger operators).
Since $P$ is of finite rank, this property follows from \eqref{e:FGR0'}.

The upper and the lower bounds of \eqref{e:FGR0} would be
crucial in our analysis. We shall keep track of the lower bound in the
sequel so as to emphasis the gain of positivity it occurs. We set
few notations.  
\begin{eqnarray}\label{e:Rbar}
R_\varepsilon:=\big((H_0-k_{i_0})^2+\varepsilon^2\big)^{-1/2},\,
\overline{R_\varepsilon}:=\overline{P} R_\varepsilon \mbox{ and }
F_\varepsilon := \overline{R_\varepsilon}^2. 
\end{eqnarray} 
Note that $\varepsilon R_\varepsilon^2=\im(H_0-k_{i_0} +i\varepsilon)^{-1}$
and  that $R_\varepsilon$ commutes with $P$. We get:  
\begin{eqnarray}\label{e:FGR}
(c_1/\varepsilon) P\geq
P\phi(\alpha)F_\varepsilon\phi(\alpha) 
P\geq (c_2/\varepsilon) P,  
\end{eqnarray} 
for $\varepsilon_0 > \varepsilon > 0$. It follows:
\begin{eqnarray}\label{e:norm}
\|R_\varepsilon\|= 1/\varepsilon \mbox{ and }  \|P\phi(\alpha)\overline
R_\varepsilon\|\leq c_1^{1/2}\varepsilon^{-1/2}.   
\end{eqnarray} 
As pointed out in Remark \ref{r:motiv}, we seek some more positivity for
the commutator above the energy 
$P=P_{k_i}\otimes P_\Omega$. We proceed like in \cite{BFSS} and set
\begin{eqnarray*}
B_\varepsilon :=\im (\overline{R_\varepsilon}^2\phi(\alpha)P).
\end{eqnarray*} 
It is a finite rank operator, see Lemma \ref{l:finite} for more
properties. 
Observe now that we gain some positivity
as soon as $\lambda\neq 0$:
  \begin{eqnarray}\label{e:motiv}
P[H_\lambda, i \lambda B_\varepsilon]P= \lambda^2 P \phi (\alpha)
F_\varepsilon \phi (\alpha) P \geq (c_2\lambda^2 /\varepsilon) P.
\end{eqnarray}
It is therefore natural to modify our conjugate operator. We set
\begin{eqnarray}\label{e:Ahat}
\hat A_n:=A_n+\lambda \theta B_\varepsilon, \mbox{ for }
n\in \N^*\cup\{\infty\}. 
\end{eqnarray} 
It depends
on the two parameters $\lambda\in \R$, $\varepsilon>0$ and on an extra
technical $\theta>0$. For the sake of clarity, we do not write these
extra dependences. Heuristically, the operator $A_\infty$ would give
the positivity around the threshold and the $B_\varepsilon$ would
complete it just above. We mention that $\hat A_\infty$ is maximal
symmetric and generates a semigroup of isometries,  see Lemma 
\ref{l:semiA}.    

\subsection{Main result}\label{s:main}
We prove the extended Mourre estimate over the threshold
$k_{i_0}$. This is the heart of the paper. The proof relies on the Feshbach method. 
We exploit the freedom we have so far
on $\varepsilon$ and $\theta$: set $\varepsilon:=
\varepsilon(\lambda)$ and $\theta=:\theta(\lambda)$ and suppose that
$\lambda=o(\varepsilon)$, $\varepsilon = o(\theta)$ and $\theta=o(1)$
as $\lambda$ tends to $0$. We summarize this into:
\begin{eqnarray}\label{e:size}
|\lambda|\,\, \ll\,\, \varepsilon \,\,\ll\,\, \theta\,\,\ll\,\, 1,
\mbox{ as } \lambda \mbox{  tends to } 0. 
\end{eqnarray} 
In \cite{BFSS}, this condition is more involved and the size of the
interval comes into the play. We stress that the conjugate operator
$\hat A_\infty$ depends on these three parameters. 

\begin{theorem}\label{t:msFGR}
Let $\Ic_0$ be a compact interval containing $k_{i_0}$ and no other
$k_i$. Assume the Fermi golden rule hypothesis \eqref{e:FGR0} and
\eqref{e:size} hold true. Suppose also that {\bf(I0)} and {\bf(I1a)}
are satisfied. Then, for  all open interval $\Ic\subset \Ic_0$:

{\rm i)} There are $M_\infty\geq 1$ and $\hat S_\infty$ a
$|H_\lambda|^{1/2}$-bounded operator such that $[H_\lambda, i\hat
A_\infty]= M_\infty+\hat S_\infty$ holds in the sense of forms on
$\Dc(N^{1/2})$.          

{\rm ii)} There is $\lambda_0>0$  such that the following  extended
Mourre estimate   
\begin{eqnarray}\label{e:msFGR}
M_\infty + \hat S_\infty \geq  
\, a(\lambda) E_\Ic(H_\lambda) -b(\lambda)E_{\Ic^c}(H_\lambda)\langle H_\lambda \rangle
\end{eqnarray} 
holds true in the sense of forms on $\Dc(N^{1/2})$,  for  all $\lambda
\in(0,\lambda_0)$. Here, one has $a(\lambda)= \lambda^2\theta c_2 /
5\varepsilon$ and  $b(\lambda)>0$.
  
{\rm iii)} If {\bf (I1b)} holds
true, then $H_\lambda$ has no eigenvalue in $\Ic$.  

{\rm iv)} If {\bf (I2)} holds true (and not necessarily  {\bf (I1b)}),
then $H_\lambda$ has no  eigenvalue in the interior of $\Ic$, for all
$|\lambda| \leq \lambda_0$. Moreover, one obtains the estimation of
the resolvents given in Theorem \ref{t:intro1}.  
\end{theorem} 

\begin{rem}
By taking $\theta$ and $\varepsilon$ as power of $\lambda$, one may take 
$a(\lambda)= \lambda^{1+\eta}/5$, for some $\eta>0$. We do
not reach the power $1$ as expected in Remark \ref{r:motiv}. This is
due to the non-linearity in $\lambda$ of the conjugate operator. Note
also, this is very small and then one does not expect a fast
propagation of the state, i.e.\ the eigenvalue turns
into a resonance. See for instance \cite{BFS2, HHH} for some lifetime estimates.
\end{rem} 

The proof of this theorem needs few steps and is given in Section
\ref{s:ext}. We first go into the Feshbach method and deal with
unperturbed spectral measure in Proposition \ref{p:est3}. Next, in
Proposition \ref{p:est4}, we change the spectral measure.

\subsection{The infrared decomposition}
As suggested by \eqref{e:motiv}, one expects to have to slip the
space with the projector $P$ to take advantage of this positivity. To
do so, we use the Feshbach method. As our result is local in energy,
we fix a compact interval $\Jc$ which contains the selected eigenvalue
$k_{i_0}$ and no others. We consider the Hilbert space $\Hr_{\Jc}:=
E_{\Jc}(H_0)\Hr$. Let $\Hr^{\rm{v}}_{\Jc}:=P\Hr_{\Jc}$ and
$\Hr_{\Jc}^{\overline{\rm{v}}}$ its orthogonal in $\Hr_{\Jc}$. The $\rm{v}$ 
subscript stands for  vacuum. Given $H$ bounded in
$\Hr_\Jc=\Hr^{\overline{\rm{v}}}_{\Jc}\oplus \Hr^{\rm{v}}_{\Jc}$, we
write it following this decomposition in a matricial way:   
\begin{eqnarray}\label{e:dec}
H= \left(\begin{array}{ll}
H^{\overline{\rv}\overline{\rv}}& H^{\overline{\rv} \rv}\\
H^{\rv\overline{\rv}}& H^{\rv\rv}
\end{array}\right).  
\end{eqnarray} 
We recall the Feshbach method, see \cite{BFS} and see also
\cite{DJ}[Section 3.2] for more results of this kind.  
\begin{proposition}\label{p:fesh}
Assume that $z\notin\sigma(H^{\overline{\rv}\overline{\rv}})$. We define
\begin{eqnarray*}
G_\rv(z)&:=& z \un^{\rv\rv} - H^{\rv\rv}-
H^{\rv\overline{\rv}}\big(z\un^{\overline{\rv}\overline{\rv}}- 
H^{\overline{\rv}\overline{\rv}} \big)^{-1}H^{\overline{\rv} \rv}.
\end{eqnarray*}  
Then, $z\in\sigma(H)$ if and only if $0\in \sigma\big(G_\rv(z)\big)$.
\end{proposition} 
The reader should keep in mind that $\Jc$ would be chosen slightly
bigger than the interval $\Ic$. This lost comes from the change of
spectral measure from $H_0$ to $H_\lambda$. 
The aim of the section is to show the following proposition about
$\hat S_\infty$, see \eqref{e:M&R'}.  
\begin{proposition}\label{p:est3}
Let $\Jc$ be a compact interval containing $k$ and no other $k_i$.
Suppose the Fermi golden rule \eqref{e:FGR0} and \eqref{e:size}, then one has
\begin{align}\label{e:est3} 
E_\Jc(H_0)\hat S_\infty E_\Jc(H_0)\geq (c_2\lambda^2\theta
\varepsilon^{-1}/3 -1) E_\Jc(H_0) 
\end{align} 
holds true in the sense of forms, for $\lambda$ small enough.
\end{proposition} 
We go through a series of lemmata and give the proof at the end of the section. 
The $-1$ of the r.h.s.\ seems at first sight disturbing as we seek for
some positivity. It would be balanced when we will add the
operator $M_\infty\geq 1$, see Section \ref{s:ext}. In the first place, we estimate the parts of $\hat S_\infty$.  

\begin{lemma}\label{l:commu'}
With respect to the  decomposition
\eqref{e:dec}, as $\lambda$ goes to $0$, we have  
\begin{eqnarray*}
E_\Jc(H_0)\big(\lambda\phi(a_\infty\alpha) - P\big)E_\Jc(H_0)=
\left(\begin{array}{cc}
O(\lambda)& O(\lambda)\\
O(\lambda)& -1
\end{array}\right). 
\end{eqnarray*} 
\end{lemma} 
\proof The part in $P$ follows directly from \eqref{e:H0p}.
The one in $\alpha$ results from Proposition \ref{p:ntau} and the fact that
$P\phi(a_\infty\alpha)P=0$.\qed   
 
\begin{lemma}\label{l:commu} Suppose that the Fermi golden rule \eqref{e:FGR0}
holds true. Then, the form $[H_\lambda, B_\varepsilon]$ defined on
$\Dc(H_\lambda)\times \Dc(H_\lambda)$ extends to a finite rank
operator on  $\Hr$, still denoted by $[H_\lambda, B_\varepsilon]$. As
$\lambda$  tends to $0$, we have   
\begin{eqnarray}\label{e:commua}
\quad \big\|\,[H_\lambda, \lambda\theta B_\varepsilon]\,\big\|_{\Bc(\Hr)}
\leq O(\lambda\theta\varepsilon^{-1/2}) + 
O(\lambda^2\theta\varepsilon^{-3/2}).
\end{eqnarray} 
Besides, with respect to the  decomposition \eqref{e:dec},  we have:   
\begin{eqnarray*}
E_\Jc(H_0)[H_0, \lambda\theta B_\varepsilon ]E_\Jc(H_0)=
\left(\begin{array}{cc}
0& O(\lambda\theta\varepsilon^{-1/2})\\
O(\lambda\theta\varepsilon^{-1/2}) & 0
\end{array}\right) 
\end{eqnarray*}
and 
\begin{eqnarray*}
E_\Jc(H_0)[\lambda \phi(\alpha), \lambda\theta B_\varepsilon]E_\Jc(H_0)=
\left(\begin{array}{cc}
O(\lambda^2\theta\varepsilon^{-3/2})& O(\lambda^2\theta\varepsilon^{-3/2})\\
O(\lambda^2\theta\varepsilon^{-3/2}) & \lambda^2\theta F_\varepsilon  
\end{array}\right).
\end{eqnarray*} 
\end{lemma} 
\proof We give some estimates independent of $\Jc$.
We expand the commutators, this could be justified by considering the
commutator in the form sense on $\Dc(H_\lambda)$. 
\begin{align}\nonumber
[\dG(\omega), \overline{R_\varepsilon}^2\phi(\alpha)P]&=[H_0-k_{i_0},
  \overline{R_\varepsilon}^2\phi(\alpha)P] \\\label{e:commu1}
&\hspace*{-2cm}=\overline{P}(H_0-k_{i_0})R_\varepsilon
\overline{R_\varepsilon}\phi(\alpha)P + \overline{P}R_\varepsilon
\overline{R_\varepsilon}\phi(\alpha)P(H_0-k_{i_0}) =
\overline{P}O(\varepsilon^{-1/2})P +0. 
\end{align}
Indeed, the first term derives from \eqref{e:norm} and
$\|(H_0-k_{i_0})R_\varepsilon\|=O(1)$. For the second one, note that
$(H_0-k_{i_0})P=0$.  

We turn to the second estimation and apply Proposition
\ref{p:ntau}. We get $\phi(\alpha)R_\varepsilon=\phi(\alpha) 
R_1R_1^{-1}R_\varepsilon = O(\varepsilon^{-1})$. By \eqref{e:norm}, we
have
\begin{align}\nonumber
[\phi(\alpha), \overline{R_\varepsilon}^2\phi(\alpha)P]&=
P F_\varepsilon P + \overline{P} \phi(\alpha)R_\varepsilon
 \overline{R_\varepsilon}\phi(\alpha)P+\overline{P}
 R_\varepsilon\overline{R_\varepsilon}\phi(\alpha)  P\phi(\alpha)(P +\overline{P})
\\ \label{e:commu2}
&= P F_\varepsilon P + \overline{P} O(\varepsilon^{-3/2})P+
\overline{P} O(\varepsilon^{-3/2})\overline{P}.
\end{align} 
Gathering lines \eqref{e:commu1} and \eqref{e:commu2}, we get
\eqref{e:commua}. We finish by adding $E_\Jc(H_0)$. \qed

We go into the Feshbach method and conclude.
\proof[Proof of Proposition \ref{p:est3}] We set
$C_\lambda:=E_\Jc(H_0)\hat S_\infty E_\Jc(H_0)$. First observe
that for all $\mu\leq -3/4$, we get
$C_\lambda^{\rm{\overline{vv}}}- \mu$ is invertible in
$\Bc(\Hr^{\overline{\rm{vv}}})$ and 
$\|(C_\lambda^{\rm{\overline{vv}}}- \mu)^{-1}\|_{\Bc(\Hr^{\overline{\rm{vv}}})}
\leq 2$. Indeed, from Lemma \ref{l:commu'} and \ref{l:commu}, we have that
  $C_\lambda^{\rm{\overline{vv}}}$ is bounded from below by 
  $O(\lambda^2\theta \varepsilon^{-3/2})+O(\lambda)$. 
This is bigger than $-1/2$ by \eqref{e:size}, for $\lambda$ small enough.

We now estimate from below the internal energy of $C_\lambda$,
uniformly in $\mu\leq 3/4$. By Lemmata \ref{l:commu'} and
\ref{l:commu}, the first part and the Fermi golden Rule \eqref{e:FGR},
we infer 
\begin{align*}
C_\lambda^{\rv\rv} -  C_\lambda^{\rm{v\overline{v}}}
(C_\lambda^{\rm{\overline{vv}}}- \mu)^{-1} C_\lambda^{\rm{\overline{v}v}}+1
\geq
&
\\
& \hspace*{-5cm} \geq c_2\lambda^2 \theta\varepsilon^{-1}+
\big(O(\lambda\theta\varepsilon^{-1/2})+ 
O(\lambda^2\theta \varepsilon^{-3/2})+ O(\lambda)\big)^2
\\
& \hspace*{-5cm} = c_2\lambda^2 \theta\varepsilon^{-1}
\big(O(\theta)+ O(\lambda \theta\varepsilon^{-1}) 
+O(\varepsilon^{1/2}) + O(\lambda^2 \theta\varepsilon^{-2}) 
+ O(\lambda \varepsilon^{-1/2}) +
O(\theta^{-1}\varepsilon )\big)  
\\
& \hspace*{-5cm} 
 \geq c_2\lambda^2\varepsilon^{-1}/2, \mbox{ for } \lambda \mbox{
   small enough.}
\end{align*}
We have used \eqref{e:size} for the last line. 

We are now able to conclude. Since $\Jc$ contains $k_{i_0}$ and no other
$k_i$. We have $E_\Jc(H_0)P_\Omega=P$ by \eqref{e:H0p}. Let $\mu<
c_2\lambda^2\theta \varepsilon^{-1}/2 -1$. Note that $\mu \leq -3/4$
for $\lambda$ small enough by \eqref{e:size}. Thanks to the previous
lower bound, we can apply Proposition \ref{p:fesh} with respect to
the decomposition \eqref{e:dec} for $C_\lambda$ and with $z=\mu$ to
get the result. \qed 

\subsection{The extended Mourre estimate}\label{s:ext}
At the end of the section, we establish the extended Mourre estimate. 
We start by enhancing Proposition \ref{p:est3}.
\begin{proposition}\label{p:est4}
Let $\Ic$ be a compact interval containing $k_{i_0}$ and not
other $k_i$.  Assume the Fermi golden rule \eqref{e:FGR0} and
\eqref{e:size}. Then, 
\begin{eqnarray*}
E_\Ic(H_\lambda)\hat S_\infty E_\Ic(H_\lambda)\geq c_2(\lambda^2\theta\varepsilon^{-1}/4
  -1) E_\Ic(H_\lambda) 
\end{eqnarray*} 
holds true in the sense of forms for $\lambda$ small enough.
\end{proposition} 
\proof
Let $\Jc$ be a compact interval as in Proposition \ref{p:est3} such
that $\Ic$ is included in its interior and contains $k_{i_0}$. By
\eqref{e:size}, it is enough to prove
\begin{eqnarray}\nonumber
 E_\Ic(H_\lambda)\big(\lambda\phi(a_\infty\alpha)+
  [H_\lambda, i\lambda\theta   B_\varepsilon]-
  P_\Omega\big)E_\Ic(H_\lambda)&\geq&\\ 
\label{e:est4'}
&&\hspace*{-6cm}
\big(c_2 \lambda^2\theta\varepsilon^{-1}/3+O(\lambda^2)+ O(\lambda^2\theta
\varepsilon^{-1/2}) + O(\lambda^3\theta \varepsilon^{-3/2})-1\big)
E_\Ic(H_\lambda).     
\end{eqnarray} 
We start with the left hand side of \eqref{e:est4'} and introduce
$E_{\Jc}(H_0)+E_{\Jc^c}(H_0)$  on the right and on the left of
$([H_\lambda, i\lambda\theta   B_\varepsilon]
+\lambda\phi(a_\infty\alpha)- P_\Omega)$. Note that both of
spectral measures are bounded in $\Dc(H_0)$, endowed with the graph
norm. We need to control the mixed term. Using Lemma \ref{l:HS}
and \eqref{e:commua}, we get  
\begin{align*}
 E_\Ic(H_\lambda)E_{\Jc^c}(H_0)[H_\lambda, i\lambda\theta B_\varepsilon ]
 E_{\Jc}(H_0)E_\Ic(H_\lambda) =&\\ 
&\hspace*{-2cm}  \big(O(\lambda^2\theta \varepsilon^{-1/2}) +
 O(\lambda^3\theta  \varepsilon^{-3/2})\big)  E_\Ic(H_\lambda),  
\end{align*} 
and a better term for $ E_\Ic(H_\lambda)E_{\Jc^c}(H_0)[H_\lambda,
 i\lambda\theta B_\varepsilon ]
 E_{\Jc^c}(H_0)E_\Ic(H_\lambda)$. Since the term
 $\phi(a_\infty\alpha)\langle H_0 \rangle^{-1/2}$ 
 is bounded in $\Hr$ by Proposition \ref{p:ntau}, Lemma \ref{l:HS} gives
\begin{eqnarray*}
 E_\Ic(H_\lambda)E_{\Jc^c}(H_0)\lambda\phi(a_\infty\alpha)
 E_{\Jc}(H_0)E_\Ic(H_\lambda) =  O(\lambda^2)E_\Ic(H_\lambda),
\end{eqnarray*} 
 and a better term for the full-mixed term. As $H_0$ commute with
 $P_\Omega$, we infer 
$ E_\Ic(H_\lambda)E_{\Jc^c}(H_0) P_\Omega E_{\Jc}(H_0)E_\Ic(H_\lambda) = 0$.
Now using Proposition \eqref{p:est3} we obtain
\begin{eqnarray*}
 E_\Ic(H_\lambda)\big([H_\lambda, i\lambda\theta B_\varepsilon ]
+\lambda\phi(a_\infty\alpha)-
P_\Omega\big)E_\Ic(H_\lambda)&\geq&\\ 
&&\hspace*{-4cm} 
(c_2 \lambda^2\theta\varepsilon^{-1}/3
-1)E_\Ic(H_\lambda)E_\Jc(H_0)E_\Ic(H_\lambda)\\
&&\hspace*{-4cm}+\big(O(\lambda^2) + O(\lambda^2\theta \varepsilon^{-1/2})
+ O(\lambda^3\theta \varepsilon^{-3/2})\big)  E_\Ic(H_\lambda).  
\end{eqnarray*}
Finally, the estimation \eqref{e:est4'} follows by noticing that
$E_\Ic(H_\lambda)E_\Jc(H_0) E_\Ic(H_\lambda)$ is equal to
$\big(1+O(\lambda^2)\big)E_\Ic(H_\lambda)$, again by Lemma
\ref{l:HS}.\qed

We are now able to prove the announced result. 
\proof[Proof of Theorem \ref{t:msFGR}]
The operator $M_\infty$ and $\hat S_\infty$ are given in \eqref{e:M&R}
and \eqref{e:M&R'}. Points i) and ii) are given in Section \ref{s:1comm'}.
By Proposition \ref{p:est4} and since $M_\infty\geq 1$, 
\begin{eqnarray*}
M_\infty + E_\Ic(H_\lambda)\hat S_\infty E_\Ic(H_\lambda)\geq
c_2\lambda^2\theta\varepsilon^{-1}/4\, E_\Ic(H_\lambda) 
\end{eqnarray*} 
holds true in the form sense on $\Dc(N^{1/2})$. Then, \eqref{e:eta} gives iii). 
The point iv) follows from the Virial Theorem, Proposition
\ref{p:6.5}. Finally Theorem \ref{t:LAP} gives point v). Indeed,
the space $\Gr$ appearing therein is identified in
Lemma \ref{l:6.4}. In remains to notice that 
the spaces \eqref{e:Gs} given for $\hat
A_\infty$ and $A_\infty$ are the same. This follows from the fact that
these operators have the same domain in  $\Gr^*$, by Lemma
\ref{l:finite} and that the  spaces $\Gr^*_s$ are given by complex
interpolation. \qed

\appendix
\section{Level shift operator}\label{s:level} 
\renewcommand{\theequation}{A.\arabic{equation}}
\setcounter{equation}{0}
In this paper, we never make the hypothesis that we analyse
an eigenvalue which  could be different than the ground state energy of
$H_0$. The point is that it is well known that  it is supposed to
remain, even if the perturbation is switched on, see for instance
\cite{AH, BFS, G0}. This leads to a contradiction to the hypothesis
made on the Fermi golden rule. Therefore, in this section, we explain
how one may check the Fermi golden rule assumption \eqref{e:FGR0},
why it is not fulfilled at ground state energy. This would also explain the compatibility
with {\bf (I0)}--{\bf (I2)}. The computations we
lead are standard, we keep it simple. See also \cite{BFS, DJ2, JP}.

Let $e_i$ be an
orthonormal basis of eigenvectors of $K$ relative to the eigenvalue
$k_i$. To simplify the computation, say that $k_{i_0}$ is
simple. Since $k_{i_0}$ is simple and since
$\phi(\alpha)(e_{i_0}\otimes \Omega)= \alpha(e_{i_0})\in \Kr\!\otimes
\hg$, \eqref{e:FGR0} is equivalent to:    
\begin{eqnarray*}
 c_1 \geq \big\langle \alpha(e_{i_0}), \im(H_0-k_{i_0}
 +i\varepsilon)^{-1}\alpha(e_{i_0}) \big\rangle \geq c_2>0, \mbox{ for } 0<
 \varepsilon \leq \varepsilon_0. 
\end{eqnarray*} 
We have $\alpha(e_{i_0})= \sum_{i=1,\ldots, n} e_i \otimes f_{i,
i_0}\in \Kr\!\otimes \hg$,  where $f_{i, i_0}=\big\langle  e_i \otimes
\un_{\hg}, \alpha(e_{i_0})\big\rangle$.  As $\hg=L^2(\R^d,dk)$, we
write $f_{i, i_0}$ as a function of $k$. We go into polar coordinates, see \eqref{e:T} and infer
\begin{eqnarray*}
c_1 \geq \sum_{i=1,\ldots, n} \int_0^\infty \int_{S^{d-1}}
\varepsilon \frac{|f_{i, i_0}|^2(r\theta) r^{d-1}} 	{(r
  +\lambda_i-\lambda_{i_0} )^2+\varepsilon^2}  \, d\sigma \, dr
\geq c_2>0 
\end{eqnarray*}
Suppose now that $(r, \theta)\mapsto |f_{i, i_0}|^2(r\theta) r^{d-1}$
is continuous and in $L^1$.  Then by dominated convergence, we let
$\varepsilon$ go to zero and get: 
\begin{eqnarray}\label{e:FGR'}
\quad\quad 	c_1 \geq \sum_{i=1,\ldots, i_0} c_i
(\lambda_{i_0}-\lambda_i)^{d-1} \int_{S^{d-1}}  |f_{i,
  i_0}|^2\big(\theta (\lambda_{i_0}-\lambda_i)\big) \, d\sigma   \geq
c_2>0  
\end{eqnarray}
Here note that, up to the constant $c_i$, $r\mapsto
\varepsilon/\big((r +\lambda_i-\lambda_{i_0})^2+\varepsilon^2\big)$
is a Dirac sequence if and only if $\lambda_i\leq\lambda_{i_0}$. 

To satisfy the Fermi golden rule, it is enough to have a non-zero term in \eqref{e:FGR'}.  
When $d\geq 2$, we stress that
the sum is taken till $i_0 -1$ and therefore  is empty at ground state
energy. When the $1$-particle space is over $\R$, it cannot be
satisfied at this level of energy as well. Indeed, one would obtain a
contradiction with the hypothesis {\bf{(I0)}} and the continuity of
$(r, \theta)\mapsto |f_{i,   i_0}|^2(r\theta)$.

\section{Properties of $C_0$-semigroups}\label{s:semiprop}
In this section, we gather various facts about $C_0$-semigroups we use
along this article.  Let $\Hr$ be a Hilbert space. Recall that  $\wlim$ denotes
the weak limit.  
\begin{definition}\label{d:semi} 
We say $\R^+\ni t\mapsto W_t$, with $W_t\in\Bc(\Hr)$ is a
\emph{$C_0$-semigroup} if 
\begin{enumerate}
	\item $W_0=\id$ and $W_{s+t}=W_sW_t$, for all $s,t\geq 0$,
	\item $\wlim_{t\rightarrow 0^+} W_t=\id$.
\end{enumerate}
\end{definition}
Automatically, this implies that $\R^+\supset t\mapsto W_t$ is
strongly continuous, see \cite{HP}[Theorem 10.6.5]. We keep the
convention of \cite{GGM2} and define the generator of $\{W_t\}_{t\geq
0}$ as being the closed densely defined operator $A$ defined on  
\begin{eqnarray*}
\Dc(A):=\{u\in\Hr \mid  \lim_{t\rightarrow 0^+}(it)^{-1}(W_t-\id)u
\mbox{ exists} \}. 
\end{eqnarray*}
We set $Au$ this limit. Formally, one reads $W_t=e^{itA}$. The map
$\R^+\supset t\mapsto W_t^*$ being weakly continuous,
$\{W_t^*\}_{t\geq 0}$ is also a $C_0$-semigroup. Its generator is
$-A^*$. We recall the \emph{Nelson Lemma}, see for instance
\cite{BR}[Corollary 3.1.7]. 
\begin{lemma}[Nelson Lemma]\label{l:Nelson}
Let $\Dc$ be a dense subset of $\Hr$ and let $\{W_t\}_{t\geq 0}$ be a
$C_0$-semigroup. If $W_t \Dc\subset \Dc$ then $\Dc$ is a core for the
generator of $\{W_t\}_{t\geq 0}$. 
\end{lemma}

Let $\Gr$ and $\Hr$ be two Hilbert spaces such that $\Gr\subset \Hr$
continuously and densely. Using the Riesz isomorphism, we identify
$\Hr$ with $\Hr^*$, where the latter is the set of  anti-linear forms
acting on $\Hr$. We infer the following scale of spaces 
$\Gr\subset \Hr \simeq \Hr^* \subset \Gr^*$ with continuous and dense
embeddings. In order to define the restriction of $W_t$ on $\Gr$, we
set: 

\begin{definition}\label{d:bstable} Given a $C_0$-semigroup
$\{W_t\}_{t\geq 0}$ on $\Hr$. We say that  $\Gr$ is \emph{b-stable}
(boundedly stable) under the action of $\{W_t\}_{t\geq 0}$ if  

{\rm i)} $W_t\Gr\subset \Gr$, for all $t\in \R^+$,

{\rm ii)} $\sup_{t\in [0,1]}\|W_t u\|$ is bounded for all $u\in\Gr$.
\end{definition}

\begin{rem}\label{r:bstable}
Note that  unlike for $C_0$-groups, the second condition is required to
ensure the continuity in $0$. These two conditions are equivalent to
the fact that $\{W_t |_{\Gr}\}_{t\geq 0}$ is a $C_0$-semigroup on $\Gr$. 
\end{rem}

 Assuming that $\Gr$ is b-stable under the action of $\{W_t\}_{t\geq 0}$, we
denote by $A_{\Gr}$ its generator. Thus,  $A_\Gr$ is the
restriction of $A$ and its domain is given by
\begin{eqnarray*}
\Dc(A_{\Gr})=\{u\in \Gr\cap \Dc(A) \mid Au\in\Gr \}.	
\end{eqnarray*}
If $\Gr^*$ is also b-stable under $\{W_t^*\}_{t\geq 0}$, we denote by
 $A_{\Gr^*}$ the generator of $\{W_t\}_{t\geq 0}$ extended to
 $\Gr^*$. As above $A$ is a restriction of $A_{\Gr^*}$ and thanks to
 the Nelson lemma, we have that $A$ is the closure of $A_\Gr$ in $\Hr$
 and that   $A_{\Gr^*}$ is the closure of $A$ in $\Gr^*$. We would drop the
 subscript $\Gr$ when no confusion could arise. 
 
We recall the following result of perturbation, see 
\cite{kato}[Theorem IX.2.1].

\begin{proposition}\label{p:kato}
Let $B$ be a bounded operator in a Hilbert space $\Hr$. Then 
$A$ is the generator a $C_0$-semigroup if and only if $A+B$ is also one.
\end{proposition} 

\section{The Mourre method}\label{s:ms}
\renewcommand{\theequation}{C.\arabic{equation}}
\setcounter{equation}{0}
\subsection{The $C^1$ class.}\label{s:C1}
Given a self-adjoint operator $A$, the so-called $C^1(A)$ class of
regularity is a key notion within the Mourre's theory, see \cite{ABG} and
\cite{GG}. This guarantees some properties of domains and that the commutator of an operator
$H$ with $A$ would be $H$-bounded.
In this paper, we have to deal with maximal symmetric
conjugate operators and thus have to extend the standard class exposed
in details in \cite{ABG}[Section 6.2]. As some refinements appear,  
we present an overview of the properties and refer to
\cite{GGM}[Section 2] for proofs. 

Within this section, we consider a closed densely defined
operator $A$ acting in a Hilbert space $\Hr$. Note this implies that
$\Dc(A^*)$ is dense in $\Hr$. We first defined the class
of bounded operators belonging to $\Cc^1(A)$. Let $S\in\Bc(\Hr)$. We
denote by $[S,A]$ the sesquilinear form defined on
$\Dc(A^*)\times\Dc(A)$ by 
\begin{eqnarray*}
\langle u, [S,A] v\rangle := \langle A^* u, S v\rangle - \langle
S^*u, Av\rangle, \mbox{ for } u\in\Dc(A^*), v\in\Dc(A).
\end{eqnarray*} 
\begin{definition}\label{d:C1bd}
An operator $S\in\Bc(\Hr)$ belongs to $\Cc^1(A)$ if the sesquilinear
form $[S,A]$ is continuous for the topology of $\Hr\times\Hr$. We
denote by $[S,A]^\circ$ the unique bounded operator in $\Hr$ extending
this form. 
\end{definition} 
We now extend the definition to unbounded operator by asking the 
resolvent  $R(z):=(S-z)^{-1}$ to be $\Cc^1(A)$. We precise the
statement. We first recall that given  $S$ a closed densely defined operator on $\Hr$, the
\emph{$A$-regular resolvent set} of $S$ is the set $\rho(S,A)\subset
\C\setminus\sigma(S)$ such that $R(z)$ is of class $\Cc^1(A)$.  

\begin{definition}
Let $S$ be a closed and densely defined operator on $\Hr$. We say
that \emph{$S$ is of class $\Cc^1(A)$} if there are a constant $C$ and a sequence of 
complex numbers $z_\nu\in\rho(S,A)$ such that
$|z_\nu|\rightarrow \infty$ and $\|R(z_\nu)\|\leq C|z_\nu|^{-1}$. If 
$S$ is of class $\Cc^1(A)$ and $\rho(S,A)=\C\setminus \sigma(S)$ then
we say that $S$ is of \emph{full class $\Cc^1(A)$}.  
\end{definition} 

In many cases these two definitions coincide. Indeed, given $S\in\Cc^1(A)$, 
one shows that if $A$ is regular
or if $S$ is self-adjoint with a spectral gap then $S$ is in the full
class $\Cc^1(A)$. We recall that a closed densely defined operator $B$ is \emph{regular}
if there is a constant $C$ and $\alpha_n\in\C\setminus \sigma(B)$ such
that 
$\|(B-\alpha_n)\|\leq C|\alpha_n|^{-1}$ and such that $|\alpha_n|\rightarrow \infty$.  The generators of $C_0$-semigroups are regular for instance.

\begin{definition}
Let $A$ and $S$ be two closed and densely defined operators in
$\Hr$. We define $[A,S]$ as the sesquilinear form acting on
$\big(\Dc(A^*)\cap \Dc(S^*)\big) \times \big(\Dc(A)\cap \Dc(S)\big)$
and given by $\langle u, [S,A] v\rangle := \langle A^* u, Sv\rangle - \langle
S^*u, A v\rangle.$
\end{definition} 
\begin{proposition}\label{p:ext}
Let $S\in\Cc^1(A)$. Then $\Dc(A^*)\cap \Dc(S^*)$ and $\Dc(A)\cap
\Dc(S)$ are cores for $S$ and $S^*$ respectively and the form $[A,S]$
has a unique extension to a continuous sesquilinear form denoted by
$[A,S]^\circ$ on $\Dc(S^*)\cap \Dc(S)$. Moreover,
\begin{eqnarray*}
[A, R(z)]^\circ=-R(z)[A,S]^\circ R(s), \mbox{ for all } z\in\rho(S,A), 
\end{eqnarray*} 
where on the right hand side, $[A,S]^\circ$ is considered as an element
of $\Bc\big(\Dc(S), \Dc(S^*)\big)$. 
\end{proposition} 
We stress the fact that $[A,S]$ extends to an element of $\Bc\big(\Dc(S),
\Dc(S^*)\big)$ is not enough to ensure $S\in\Cc^1(A)$, see
\cite{GG}. Some conditions of compatibilities are to be added, see
\cite{GGM}[Proposition 2.21]. This
could also be bypassed by knowing some invariance under a
$C_0$-semigroup generated by $A$.

\begin{definition}
Let $\{W_{1,t}\}_{t\in\R^+}, \{W_{2,t}\}_{t\in\R^+}$ be two
$C_0$-semigroups on the Hilbert spaces $\Hr_1$ and $\Hr_2$ with
generator $A_1$ and $A_2$. We say that $B\in\Bc(\Hr_1, \Hr_2)$ is of 
\emph{class $\Cc^1(A_1, A_2)$} if:
\begin{eqnarray*}
 \|W_{2,t}S- SW_{1,t} \|_{\Bc(\Hr_1, \Hr_2)}\leq ct, \, 0\leq t\leq 1.
\end{eqnarray*} 
\end{definition} 
If $\Gr\subset \Hr$ are two Hilbert spaces continuously and densely
embedded and if a $C_0$-semigroup $\{W_t\}_{t\in\R^+}$, with generator
$A$ on $\Hr$, $b$-stabilizes $\Gr$ and $\Gr^*$, we denote the class
$\Cc^1(A_\Gr, A_{\Gr^*})$ by $\Cc^1(A;\Gr, \Gr^*)$. We have the
following result.
\begin{proposition}\label{p:2comm}
$S\in\Cc^1(A_1, A_2)$ if and only if the sesquilinear form
${}_2[S,A]_1$ on $\Dc(A_2^*)\times \Dc(A_1)$ defined by $\langle
u_2, {}_2[S,A]_1 u_1\rangle:= \langle S^* u_2, A_1 u_1\rangle- \langle
A^*_2u_2, S u_1\rangle $ is bounded for the topology of $\Hr_2\times
\Hr_1$. Let ${}_2[S,A]_1^\circ$ be the closure of this form in
$\Bc(\Hr_1, \Hr_2)$. We have:
\begin{eqnarray*}
 {}_2[S,A]_1^\circ=\slim_{t\rightarrow 0^+}(S W_{1,t}- W_{2,t}S). 
\end{eqnarray*} 
\end{proposition} 
Note that for $S\in \Bc(\Hr)$, with $\Hr_i=\Hr$ and $W_{i,t}=W_t$, one has
$S\in \Cc^1(A_1, A_2)$ if and only if $B\in\Cc^1(A)$.

\subsection{Regularity assumptions for the limiting absorption
principle}\label{s:LAPhyp} 
In this part, we recall a set of assumptions presented in \cite{GGM}
so as to ensure a limiting absorption principle, see Theorem
\ref{t:LAP}. Consider $H$ a self-adjoint operator, $H'$ symmetric
closed and densely defined and $A$ closed and densely defined. These
operators are linked by $H'=[H,iA]$ in a sense defined lower. Denote
also $\Dr:=\Dc(H)\cap \Dc(H')$ endowed with the intersection topology,
namely the topology associated to the norm $\|\cdot\|+ \|H
\cdot\|+\|H'\cdot\|$.   

We start by some assumptions on $H$ and on $H'$.
\begin{itemize}
\item[{\bf (M1)}] \emph{$H$ is of full class $\Cc^1(H')$,
$\Dr=\Dc(H)\cap \Dc(H'^*)$ and this is a core for $H'$.} 
\item[{\bf (M2)}] \emph{There are $\Ic\subset\R$ open and
bounded and $a,b\pg 0$ such that} 
\begin{equation}\label{e:mstrict}
H'\geq \big(a \un_\Ic(H) -b \un_{\Ic^c}(H) \big)\langle
H\rangle
\end{equation}
\emph{holds true in the sense of forms on $\Dr$.} 
\end{itemize}

The last one is the \emph{strict Mourre estimate}. In order to check
the first hypothesis, we rely on \cite{GGM}[Lemma 2.26], see also
\cite{Ski}[Lemma 2.6]:    
\begin{lemma}\label{l:avoid}
Let $H,M$ be self-adjoint operators such that $H\in\Cc^1(M)$ and that
$\Dc(H)\cap \Dc(M)$ is a core of $M$. Let $R$ be a symmetric operator
such that $\Dc(R)\supset \Dc(H)$. Set $H'$ the closure of $M+R$
defined on $\Dc(R)\cap\Dc(M)$. Then $H$ is of full class $\Cc^1(H')$
and  $\Dc(H)\cap\Dc(H')$ is a core for $H'$ and $\Dc(H)\cap \Dc(H')=
\Dc(H)\cap \Dc(H'^*)=\Dc(H)\cap \Dc(M)$. 
\end{lemma}
Assuming {\bf (M2)}, one chooses $c\pg 0$ such that  $H'+c \langle
H\rangle \geq \langle H\rangle$ (take for instance $c=b+1$). Since
$H'+c \langle H\rangle$ is symmetric and positive, it possesses a
Friedrichs extension $G\geq \langle H\rangle$. We name the form domain
of $G$: 
\begin{eqnarray}\label{e:G}
\Gr:=\Dc(G^{1/2}), \mbox{ endowed with the graph norm } \|\cdot\|_\Gr.	
\end{eqnarray} 
Note that $\Gr$ is also obtained by completing the space $\Dr$ with
the help of the norm $\|u\|_\Gr= \sqrt{\langle u, (H'+c \langle
H\rangle)u \rangle}$. We identify these spaces in Lemma \ref{l:6.4}.

We now recall the dual norm $\|\cdot\|_{\Gr^*}$ of $\Gr$. Given
$u\in\Hr$, we set 
\begin{eqnarray}\label{e:dual}
\|u\|_{\Gr^*}:=\sup_{v\in\Dr, \, \|v\|_{\Gr}\leq 1} |\langle u,
v\rangle|= \|G^{-1/2}u\|. 
\end{eqnarray}
Using the Riesz isomorphism, we identify $\Hr$ with $\Hr^*$ the space
of anti-linear forms on $\Hr$. The space $\Gr^*$ is given by the
completion of $\Hr$ with respect to the norm $\|\cdot\|_{\Gr^*}$. We
get the following scale space: 
\begin{eqnarray*}
\Dr\, \subset \, \Gr \, \subset \, \Hr\simeq \Hr^*\, \subset \, \Gr^* \, \subset \, \Dr^*,	
\end{eqnarray*}
with dense and continuous embeddings. 

We turn to the assumptions concerning the conjugate operator $A$ and
higher commutators. Suppose $A$ to be the generator of $\{W_t\}_{t\in\R^+}$
\begin{itemize}
\item[{\bf (M3)}] \emph{The $C_0$-semigroup $\{W_t\}_{t\in\R^+}$ is
  of isometries and  $b$-stabilizes $\Gr$ and $\Gr^*$,} 
\item[{\bf (M4)}] \emph{$H\in\Cc^1(A; \Gr, \Gr^*)$,}
 \item[{\bf (M5)}] \emph{$H\in\Cc^{1,1}(A; \Gr, \Gr^*)$.}
\end{itemize}
The hypothesis {\bf (M4)} implies that 
\begin{eqnarray*}
 \lim_{t\rightarrow
0^+}\big(\langle u, W_t Hu\rangle -  \langle Hu, W_t u\rangle\big)= 
\langle u, H'u\rangle, \mbox{ for all } u\in\Dr.
\end{eqnarray*} 
The hypothesis {\bf (M5)} means that  $H\in \Bc(\Gr, \Gr^*)$ and  that
\begin{eqnarray*}
\int_0^1 \big\|\, [W_t, [W_t, H]]\, \big\|_{\Bc(\Gr, \Gr^*)}\, \frac{dt}{t^2} <\infty.
\end{eqnarray*} 
This is equivalent to the fact that $H$ belongs to $\big(\Cc^{2}(A; \Gr,
\Gr^*), \Bc(\Gr, \Gr^*)\big)_{1/2,   1}$. We refer to \cite{ABG, Tri} for
real interpolation.  

One may also consider the stronger $H'\in\Cc^1(A; \Gr, \Gr^*)$, i.e.\
\begin{itemize}
\item[{\bf (M5')}] \emph{$H\in\Cc^{2}(A; \Gr, \Gr^*)$.}
\end{itemize}

We now give the result. Let $A_{\Gr^*}$ be the generator of
$\{W_t\}_{t\in\R^+}$ generator in $\Gr^*$. For $s\in(0,1)$, we set:
\begin{eqnarray}\label{e:Gs}
\Gr^*_s:=\Dc\big(|A_{\Gr^*}|^s\big)  \mbox{ and } \Gr_{-s}:=(\Gr^*_s)^*.	
\end{eqnarray} 
Here, the absolute value is taken with respect to the Hilbert
structure of $\Gr^*$. Given $\Jc$ an interval, we define
$J_0^{\pm}:=\{\lambda\pm i \mu, \lambda\in \Jc \mbox{ and } \mu >0\}$.
Finally, set $R(z):=(H-z)^{-1}$. From \cite{GGM}, we obtain:
\begin{theorem}\label{t:LAP}
Assume that {\bf(M1)}--{\bf(M5)} hold true. Let $\Jc$ be a compact
interval included in $\Ic$. Then if $z\in \Jc_0^\pm$, $R(z)$ induces a
bounded operator in $\Bc(\Gr^*_s, \Gr_{-s})$, for all $s\in (1/2,
1]$. Moreover the limit $R(\lambda\pm i0)= \lim_{\mu\rightarrow \pm 0}
R(\lambda+i\mu)$ exists in the norm topology of $\Bc(\Gr^*_s,
\Gr_{-s})$, locally uniformly in $\lambda\in \Jc$ and the maps
$\lambda\mapsto R(\lambda \pm i0)\in\Bc(\Gr^*_s, \Gr_{-s}) $ are
H\"older continuous of order $s-1/2$. 
\end{theorem}
This theorem can be improved by considering weights in some Besov
spaces related to the conjugate operator. We refer to \cite{GGM} for more
details. Note that the theory exposed in \cite{GGM} is formulated 
with the hypothesis {\bf (M5')} but, as mentioned in \cite{GGM} and
proceeding like in \cite{ABG} for instance, the hypothesis {\bf (M5)}
is enough to apply the theory.

\bibliographystyle{alpha}
\def\cprime{$'$} \def\cprime{$'$} \def\cprime{$'$} \def\cprime{$'$}
  \def\cprime{$'$}

\end{document}